# Aerosols and Methane in the Ice Giant Atmospheres Inferred from Spatially Resolved, Near-Infrared Spectra: I. Uranus, 2001-2007

*Michael T. Roman [1], Don Banfield [2], Peter J. Gierasch [2]

[1] Department of Astronomy, University of Michigan

[2] Department of Astronomy, Cornell University

* mtroman@umich.edu

ABSTRACT

We present a radiative transfer analysis of latitudinally resolved H (1.487-1.783 μm) and K (2.028-2.364 μm) band spectra of Uranus, from which we infer the distributions of aerosols and methane in the planet's atmosphere. Data were acquired in 2001, 2002, 2004, 2005, and 2007 using the 200-inch (5.1m) Hale Telescope and the Palomar High Angular Resolution Observer (PHARO) near-infrared adaptive optics (AO) camera system (Hayward, 2001). Observations sample a range of latitudes between ~80°S and ~60°N on the Uranian disk. At each latitude, a vertical distributions of aerosols was retrieved using a custom non-linear constrained retrieval algorithm. Two layers of aerosols are needed to match the observations: a thin upper layer peaking just below the 100-mb tropopause and a lower clouds at ~1.9 bars. Latitudinal variations in aerosols are interpreted in context of notional circulation models, while temporal changes suggest potential seasonal effects. We infer significant reduction in aerosol scattering optical thickness in southern latitudes between 2001 and 2007, in agreement with trends reported in studies covering part of the same period using different data and retrieval algorithms (e.g., Irwin et al., 2009, 2010, 2012; Sromovsky et al., 2009). Best fits to the data are consistent with proposed models of polar depletion of methane (e.g., Karkoschka and Tomasko, 2011). Finally, a discrete cloud from 2007 is analyzed in context of simple parcel theory, with the goal of identifying likely formation mechanisms. The low scattering optical thicknesses of the discrete high cloud are consistent with formation associated with vortices and shallow lift rather than deep convection.

*1. Introduction*

The atmosphere of Uranus has a history of variable cloud activity, but the processes responsible for the observed cloud distributions and their variability are not well understood. The clouds and hazes may be expected to respond to the seasonal forcing associated with Uranus' extreme axial tilt, but the ice giant's great distance, long seasonal cycle, and typically subtle atmospheric features make detailed observations of the clouds and their variability challenging.

Numerous spectroscopic and photometric measurements have led to our current knowledge of Uranus' clouds and hazes. Though a variety of cloud models have been employed, there is a consensus that two or more vertical layers of aerosols are required to reproduce the observed reflectance—including, at least, a high tenuous haze above a more substantial cloud layer—within an atmosphere of increasing methane with depth (e.g. Neff et al., 1984; Baines and Bergstralh, 1986; Pollack et al., 1986; Sromovsky and Fry, 2007; Sromovsky et al., 2011; Irwin et al., 2009; Irwin et al 2010; Irwin et al., 2012a; Tice et al., 2013; Irwin et al., 2015). With recent improvements in methane absorption coefficients, near-infrared (NIR) radiative transfer studies have been particularly fruitful in characterizing cloud vertical structure, latitudinal variation, and aerosol properties (Irwin et al., 2010, 2012, 2015).

Nevertheless, uncertainties in the aerosol concentrations and methane mole fraction remain due to inherent challenges in constraining each remotely. Both methane and aerosols are expected to vary spatially and both contribute to the reflectance over much of the observed spectrum, making it difficult separate the precise contribution of each. In the NIR studies, this problem has often been dealt with by assuming methane profiles guided by theory and limited data, while retrieving aerosol profiles consistent the observed spectra (e.g. Irwin et al. 2010, 2012, 2015); given the interdependence, uncertainties in the methane profile produce uncertainties in the retrieved aerosol profiles.

Voyager radio occultation data combined with infrared IRIS observations provided a range of possible methane and temperature profile combinations (Lindal et al, 1987). More recently, the best estimates of the stratospheric methane mole fraction have come from observations in the far infrared, where the contribution from aerosols is minimized (Orton et al., 2014; Lellouch et al., 2015). Analysis of Spitzer IRS spectra at wavelengths from 7.4-9.5 μm suggest a methane profile with a stratospheric mole fraction of $(1.6 +.2 /-.1) \times 10^{-5}$, far less than saturation limited and what is inferred in Neptune's atmosphere (Orton et al., 2014; Lellouch et al., 2015)). Subsequent analysis of Herschel/PACS and HIFI data by Lellouch et al. (2015) found a 6 times greater methane mole fraction at the 200-mbar height and suggested that the two values could be reconciled if the methane mole fraction decreases sharply from $4.7 \times 10^{-5}$ at the 89 mbar temperature minimum to match the Spitzer value at 2 mbar. As we will show, the latter methane profile would require significantly more aerosol in the haze layer to match observations.

Deeper in the atmosphere, wavelengths with varying sensitivities to methane and hydrogen absorption have been used to mitigate degeneracies. By comparing of hydrogen and methane absorption in HST-STIS spectra from 300-1000 nm, Karkoshcka and Tomasko (2009) found that the methane abundance in the 1-3 bar heights varied with latitude, with depletion towards the poles indicative of dynamical circulation (Karkoschka and Tomasko, 2011). Sromovsky et al. (2011) and Sromovsky et al. (2014)

also determined methane depletion at the poles, but with slightly greater abundance at depth. How these latitudinal variations change with time has yet to be determined.

Details of how aerosols and methane distributions may change with the seasons are only emerging as detailed observations continue to sample the atmosphere over a collectively greater fraction of the Uranian seasonal cycle. Comparing observations over several years, significant changes in aerosols can be seen between 2006 and 2011 (Irwin et al., 2009, 2010, 2011, 2012; Sromovsky et al., 2009), with a trend in diminishing reflectance at southern latitudes suggesting seasonal changes.

Against a backdrop of seasonal scale variability, occasional transient discrete cloud features appear that usually change on scales of months or less with undetermined formation dynamics (Sromovsky and Fry, 2005). Reports of occasional discrete bright cloud features on Uranus extend at least as far back as 1870, with some spots persistent enough to be observed over multiple nights and bright enough to significantly increase the perceived disk-brightness. Likewise, observers noted bright equatorial zones and adjacent dark belts, likened to those on Jupiter, beginning in the 1880s and continuing into the first half of 20th century (for an historical perspective, see Alexander, 1965 and references therein). By the 1970s, cloud activity on Uranus appears to have diminished, and the lack of attenuating aerosols in spectroscopic data had led some at the time to speculate whether or not the visible atmosphere was entirely devoid of clouds (Belton, McElroy, and Price 1971; Belton and Price, 1973). When Voyager 2 encountered Uranus is 1986, visible images of the sunlit, southern hemisphere (shortly after solstice) had captured only low-contrast features and subtle banding on an otherwise indistinct disk (Smith, et al., 1986). The few discrete cloud features seen were interpreted as high clouds, likely composed of methane, and thought to possibly be signs of localized increased convective activity (Smith, et al., 1986).

In the years since the Voyager 2 encounter, discrete cloud activity on Uranus appears to have increased, and modern technology has facilitated study of the enhanced activity. Using the Hubble Space Telescope (HST) and terrestrial telescopes fitted with adaptive optics, observers have documented a number of particularly bright features on Uranus at mid-Northern latitudes since the early 1990's (Karkoschka, 1998 & 2001; Sromovsky et al. 2000; Sromovsky and Fry, 2004; Sromovsky et al., 2007; de Pater et al., 2014). Some features at roughly 30°N latitude appear to be part of a long-lived complex that produced exceptional bright clouds in Keck observations from August 2005 and June 2007 (Sromovsky et al., 2009). The complex appeared to drift in longitude and oscillate in latitude, and it was likely associated with a dark spot seen at the same latitude seen by HST in 2006 and likely again by Keck in the 2007 observations (Sromovsky et al., 2009; Hammel et al., 2009), inevitably calling to mind the Great Dark Spot and companion clouds of Neptune as seen by Voyager 2. Observations in 2014 captured clouds of brightness and extent unprecedented in the modern era at similar latitudes (see de Pater et al., 2015). What mechanisms lead to the occasional formation of these clouds is an open question, but some potential clues into discrete cloud dynamics may be inferred from analysis of their vertical structure.

In the present study, we add to previous investigations into the distribution of Uranus' clouds and hazes by analyzing an unpublished data set of spatially resolved, near-infrared spectral data—acquired 2001 and 2007 using the Hale 200-inch telescope at the Palomar Observatory, aided with adaptive optics. Using a customized constrained inversion algorithm, vertical profiles of clouds and hazes are retrieved over a range of latitudes as several methane models are investigated. The retrieved aerosol distributions are in good agreement with previously published results (e.g. Irwin et al., 2009; Irwin et al 2010), despite differences in retrieval techniques and modeling assumptions. The consistency of the acquisition and analysis provides an ideal opportunity for analyzing potential seasonal changes in the years leading up to equinox, complimenting other temporal studies that sampled only part of the same period using a variety of data sets (e.g. Irwin et al., 2009, 2010, 2011, 2012; Sromovsky et al., 2009). Apparent seasonal trends in the clouds and hazes are reported, consistent with the findings from subsequent observations. Finally, a lone, discrete cloud feature is analyzed; with the retrieved vertical structure and optical thickness, a simple thermodynamic parcel theory model is used to argue in favor of a formation associated with shallow lift as opposed to deep convection. Section 2 describes the data set, followed by the analysis techniques in section 3, and results in section 4. Section 5 presents a discussion of the results, concluding with a summary in section 6.

## 2. Data
### 2.1. Observations

Observations of Uranus were made annually between 2001 and 2007 at Palomar Observatory, using the 200-inch (5.1m) Hale Telescope and the Palomar High Angular Resolution Observer (PHARO) near-infrared adaptive optics (AO) camera system (Hayward, 2001). The PHARO instrument captured spatially resolved images and spectra in both the H (1.487-1.783 μm) and K (2.028-2.364 μm) bands. Images had a plate scale of 40 mas per pixel, with Uranus extending ~93 pixels in diameter in the 1024

X 1024 array images. The corrected seeing discs (i.e. the point spread function through the atmosphere following the adoptive optics correction) were estimated from the full width at half maximum of Uranian satellites; depending on the atmospheric seeing and AO performance, corrected seeing disks were typically four to seven pixels in diameter near the planet center, equivalent to roughly 2160 km to 3780 km of spatial scale at the sub-observer point on Uranus—far short of being diffraction limited. The spectrograph had a resolving power of roughly R~1500. Passing through a 0.5 arc second slit (~13 pixel), the observed flux was dispersed by grating prism across a 1024-pixel detector at a resolution 0.285 and 0.332 nm/pixel for H and K, respectively. The slit of the spectrograph was aligned roughly along the planet's central meridian, yielding spectra as a function of latitude across the disc. Precise alignment was not possible due to operational constraints imposed by the AO system. The final calibrated spectra were binned by a factor of 13 to match the spectral resolution set by the spectrographs' projected slit width, yielding an effective resolution of R~115. Spectra were also averaged spatially (in latitude) with immediately adjacent pixels to improve the signal to noise just prior to analysis. Data were collected by a number of observers including Don Banfield, Phil Nicholson, Daphne Stam, and Barney Conrath among others. Additional observations were scheduled for September 2009 but were precluded on account of airborne ash from wildfires (the Station Fire) in the nearby Angeles National Forest. Annual observations then ceased as Cornell University's fraction of Palomar time diminished after 2009.

The observations cover about 1/12 of a Uranus orbit (84 years), approaching the December 2007 Uranus equinox marking the seasonal passage of the sub-solar latitude northward across the equator (see Fig 1). For reference, Voyager 2 had observed the southern hemisphere during the preceding solstice. While the present observations were made over multiple nights each year from 2001 to 2007, many nights failed to yield sufficiently reliable data due to poor AO performance, bad seeing, or sub-optimal photometric conditions; as a result, no data from 2003 and 2006 are included in the analysis (see Table 1). All H-band images used in the study are shown in Figure 2. Solar phase angles were less than 2°.

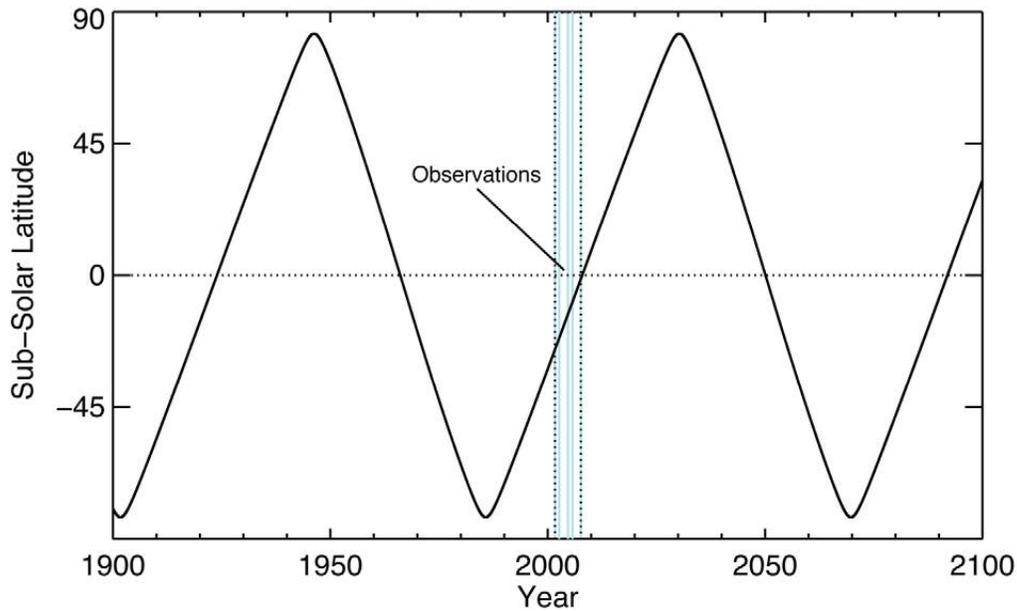

Figure 1. The Sub-solar latitude is plotted as a function of year to illustrate the seasonal coverage of the data set. Observations cover a period of six years from late southern-summer to just prior to equinox.

Table 1. Observations

| Date (year-month-day, hr:mn UTC) | Sub-Earth Latitude (degrees) | Sub-Solar Latitude (degrees) | Disk Size (arcseconds) | Solar Phase Angle (degrees) |
|---|---|---|---|---|
| 2001-09-03, 08:07 | -26.22 | -25.27 | 3.71 | 0.94 |
| 2001-09-04, 08:40 | -26.26 | -25.26 | 3.70 | 0.99 |
| 2002-08-24, 09:09 | -21.59 | -21.36 | 3.71 | 0.23 |
| 2002-08-24, 09:38 | -21.59 | -21.36 | 3.71 | 0.23 |
| 2002-08-25, 09:26 | -21.63 | -21.35 | 3.71 | 0.28 |
| 2002-08-26, 09:20 | -21.67 | -21.34 | 3.71 | 0.33 |
| 2002-08-26, 09:52 | -21.67 | -21.34 | 3.71 | 0.33 |
| 2004-07-25, 10:28 | -11.98 | -13.62 | 3.67 | 1.59 |
| 2004-07-25, 10:56 | -11.98 | -13.62 | 3.67 | 1.59 |
| 2004-07-28, 11:25 | -12.08 | -13.59 | 3.68 | 1.46 |
| 2004-07-28, 11:53 | -12.08 | -13.59 | 3.68 | 1.46 |
| 2005-09-16, 07:59 | -9.78 | -9.00 | 3.69 | 0.76 |
| 2005-09-16, 08:30 | -9.78 | -9.00 | 3.69 | 0.76 |
| 2005-09-16, 08:56 | -9.78 | -9.00 | 3.69 | 0.76 |
| 2005-09-16, 09:24 | -9.78 | -9.00 | 3.69 | 0.76 |
| 2007-08-15, 11:34 | 0.03 | -1.26 | 3.68 | 1.24 |
| 2007-08-15, 12:04 | 0.03 | -1.26 | 3.68 | 1.24 |
| 2007-08-17, 09:13 | -0.04 | -1.24 | 3.68 | 1.15 |
| 2007-08-17, 09:47 | -0.04 | -1.24 | 3.68 | 1.15 |
| 2007-08-17, 10:16 | -0.04 | -1.24 | 3.68 | 1.15 |

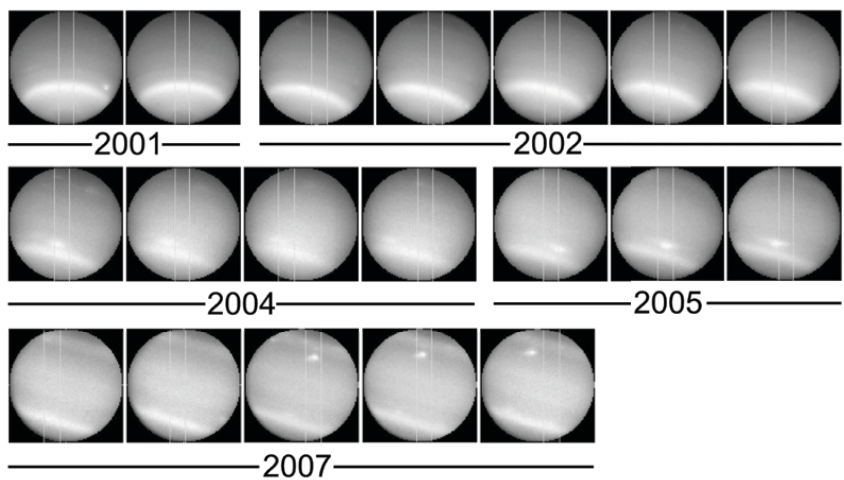

Figure 2. H-band images of Uranus, in chronological order corresponding to Table 1,. Images are calibrated but shown arbitrarily scaled to enhance contrast. Faint pairs of dotted vertical lines are added indicate the position of the spectrograph slit. Uranus' south pole is down in the images.

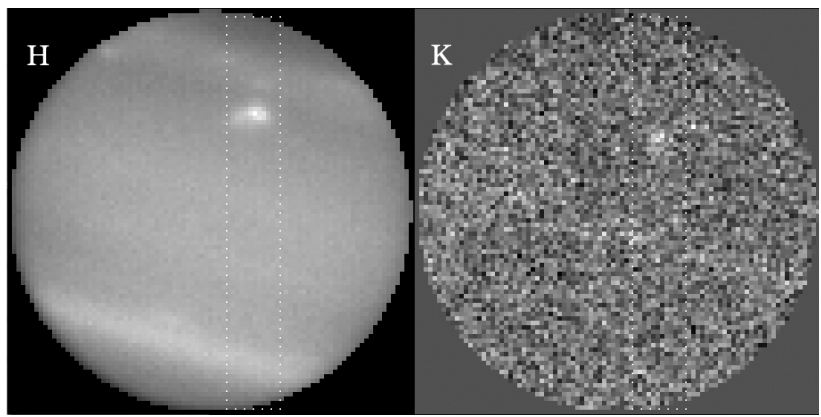

Figure 3. Example images of Uranus in H and K bands captured on August 17, 2007. Bright polar collars and cloud features can be seen in H images; one feature was seen in the K images that were otherwise dominated by noise.

Images of Uranus in the H-band showed a mostly bland disc marked with a few distinguishable features (Fig. 3). A bright band, commonly referred to as the south polar collar, was visible at latitudes roughly between 40°S and 50°S over the entire period of observations. Likewise, a bright feature was often seen at latitudes between roughly 25°S and 35°S, just equator-ward of the polar collar; others researchers have noted a similar cloud complex (informally referred to as Berg) at this latitude (Sromovsky and Fry, 2005; Sromovsky et al., 2009). Occasional discrete cloud features were seen at mid-northern latitudes. With the exception of a single feature visible at roughly 30°N in 2007, none of the aforementioned features were seen in the K-band images (Fig. 3). Due to gaseous molecular hydrogen and methane absorption, K-band images and spectra were extremely dark and dominated by noise, with only very slight signal evident at the short end of the K-band spectra; the most obvious feature of these images and spectra are the Uranian rings, exposed by the long integration times needed to increase the signal to noise ratio. One example of the spectra is shown in figure 4.

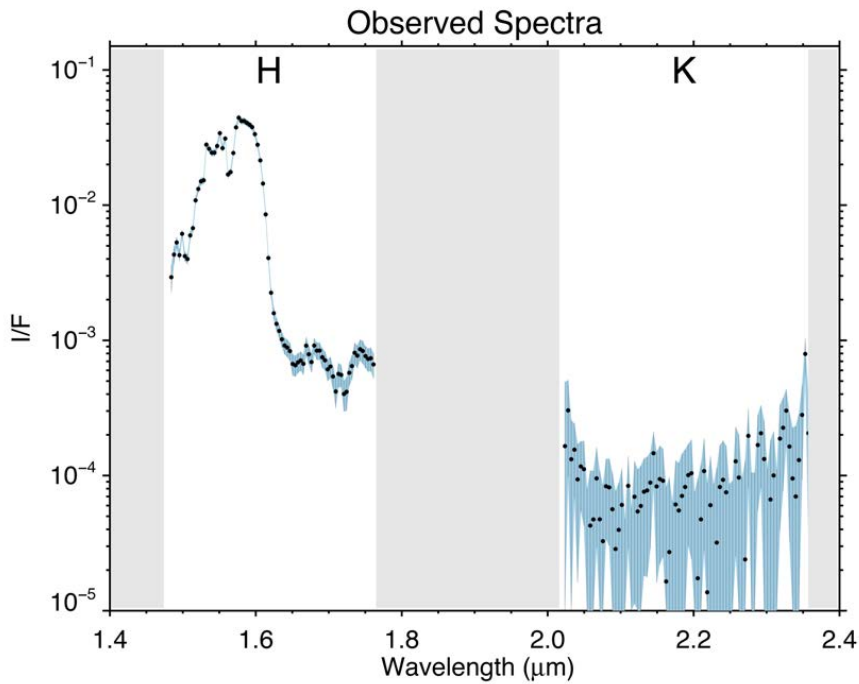

Figure 4. Example of H & K band spectra. The dots represent the calibrated, observed reflectance expressed in unitless I/F as a function of wavelength, binned to a resolution set by the spectrograph slit-width. The dark shading depicts the measurement uncertainties. Regions outside the H and K band spectral passbands are lightly shaded. This observation corresponds to 22°S latitude on September 4, 2001.

*2.2 Calibrations and Processing*

The data were reduced and photometrically calibrated using standard techniques for near-infrared spectra and images, including flat fielding, sky-subtraction, and interpolation to remove bad pixels. The faint standard star, FS34 was used for flux calibration, while a number of solar-type stars (SAO 163989, SAO 164580, SAO164338, SAO 146135, SAO 163616) were used as solar proxies, allowing us to express the calibrated spectra into dimensionless units of I/F.

Spectra were calibrated in wavelength by comparison with known spectral lines of planetary nebula NGC 7027 and carefully corrected to remove errors due to imperfect sky subtraction and occasionally detectable tilts of the slit relative to the detector.

The edges of the disk pose an additional calibration issue. Though adaptive optics largely corrects for the atmospheric seeing, residual blurring still alters the spatial distribution of the flux. Light along the edge of the planetary disk blurs with the sky beyond the disk, reducing the observed flux along the edge of the planet. Likewise, neighboring dimmer regions reduce the flux from distinctly brighter regions on the disk. In imaging studies, this blurring has previously been modeled as the convolution of the true image (i.e. the image prior to being distorted by the atmosphere) and the point spread function (PSF) of the Earth's atmosphere (Karkoschka, 2001; Sromovsky and Fry, 2007). The true image could then be largely restored by de-convolving the observed image with the correct atmospheric PSF, with the cost of amplifying the noise (e.g. see Sromovsky and Fry, 2007). Some additional smoothing is required to correct for the increased noise, typically in the form of Gaussian convolution. The correct PSF is a function of time, position, and wavelength, but it may be estimated from the image data itself if one assumes the brightness distribution of the true disk.

Deconvolving the spectra is not as simple, since the spectral observations are in one spatial dimension while the blurring occurs in two spatial dimensions; the flux lost beyond the edges of the slit cannot be recovered in the spectral images, and the 1-dimensional PSF required for a 1-d deconvolution would need to be artificially adjusted to compensate for this loss. Furthermore, the spectral deconvolution significantly amplifies the noise, and in darker regions of the spectra this increased noise can completely dominate over the signal. For these reasons, deconvolving the spectra introduces the risk of corrupting the true data. To assess the potential effect on the results, representative cases from 2001 and 2007, representing the end points of the time span, were deconvolved and analyzed. For these two cases, deconvolutions were performed only on the image observations to avoid uncertain manipulations of the spectra, and the modified image flux was then used to calibrate the spectra. The atmospheric PSF was estimated by convolving a sharp synthetic featureless disk, representing the true disk image, with an adjustable PSF (with a functional form as applied by Sromovsky and Fry, 2007); PSF parameters were adjusted until the synthetic disk matched the observed disk; this PSF was assumed to match the atmospheric PSF and hence used to deconvolve the observations. A Gaussian with a full-width at half-maximum of 4 pixels was employed in the deconvolutions to keep the noise from increasing by more than 10%. The resulting H spectra are overall brighter—by up to 30% brighter for locations corresponding to very edges and discrete bright features, but indeed noisier.

Rather than introducing additional noise or risk any spurious manipulation of these data, we chose to simply omit locations approaching the edge of the planet's disk from our latitudinal cross sections and note that scattered optical thicknesses may still be underestimated by up to 10% towards the edges. Based on the deconvolutions analysis, an emission angle cutoff of 70° was deemed sufficiently far enough from the edge to keep errors in I/F well below 10%. This corresponded to about 6 pixels from the edge. Discrete feature may still be underestimated in this approach, leading to an underestimate of the retrieved scattering optical depth of less than 10% for the brightest, discrete features.

## *3. Analysis*

### *3.1 Spectral sensitivities*

Aerosol scattering and molecular absorption of sunlight shape the spectrum of Uranus in the H and K bands. At these near-infrared wavelengths, Rayleigh and Raman scattering are insignificant, and absorption from molecular hydrogen and methane would cause the atmosphere to appear dark in the absence of aerosols.

The wavelength dependence of the molecular absorption can be seen in a plot of transmission depths computed from gaseous absorption coefficients. Figure 5 shows the atmospheric pressure heights corresponding to a single optical depth of attenuation along the two-way path for reflected sunlight at normal incidence for two different assumed methane models. The first model assumes that methane is saturation limited in the upper troposphere and thus reaches a constant minimum value set by the coldest height; in the second model (as shown in Fig. 12), the methane mole fraction is less, only $1.6 \times 10^{-5}$, consistent with values inferred from Spitzer observations (Orton, et al., 2014). Both models assume a deep methane abundance of 3.2% for pressure greater than 2 bars, following Karkoschka & Tomasko (2009), and a temperature profile of Lindal et al., 1987 (model F).

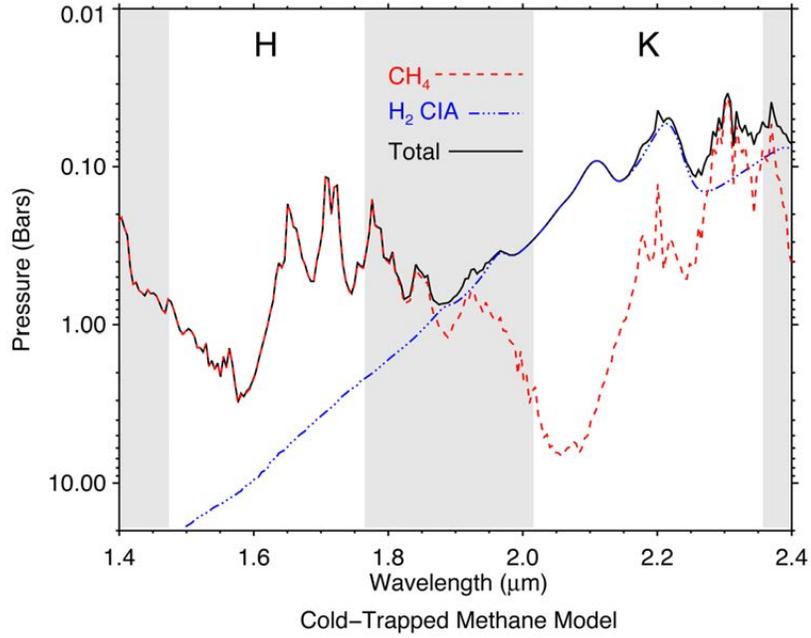

Cold−Trapped Methane Model

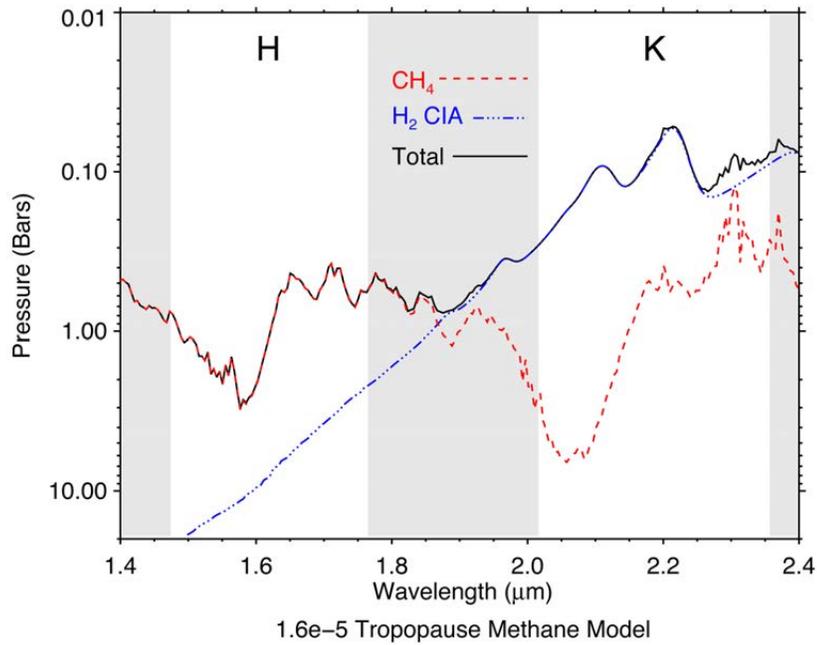

1.6e−5 Tropopause Methane Model

Figure 5. Uranus clear atmosphere penetration depths. Atmospheric pressure heights corresponding to a single optical depth of attenuation along the two-way path for reflected sunlight at normal incidence through an aerosol-free atmosphere. Different components to the attenuation are plotted separately. Weak methane absorption dominates in the H-band; collisional induced absorption by hydrogen dominates the K-band. Wavelengths beyond the filter ranges are shaded in gray.

Methane is the primary absorber in the H-passband, but the wavelength dependence of the absorption results in a wide range of vertical sensitivities. In the center of the H-band, the wavelength-dependent methane absorption is relatively weak, and incident light can penetrate and return from several bars of pressure before it is considerably attenuated. Towards the edges of the band, the absorption is greater and radiation is attenuated over a much shorter path length.

Through much of the K-band, absorption due to the collisional induced dipole of molecular hydrogen (hydrogen-hydrogen and helium-hydrogen collisions) dominates over the methane absorption. In the hydrogen-rich atmosphere of Uranus, this serves as a significant source of opacity. Optical depths of unity are reached at pressures less than a few hundred millibars, making the K-band sensitive to aerosols located above a few hundred millibars. At the longest K-band wavelengths, methane once again becomes a significant source of opacity, comparable in contribution to hydrogen CIA.

Assuming that the atmospheric mixing ratio of hydrogen is well constrained, then the atmospheric opacity in the K band (where hydrogen absorption largely dominates) may be reliably computed. Given a range of sensitivities to the hydrogen absorption across the spectrum, observations span a range of heights. By comparing the reflectance at different wavelengths, the vertical distribution of the scatterers can potentially be inferred. For K-band observations, the transmission sensitivities permit vertical resolution within a layer between 60 mbar and 300 mbar; if a cloud existed anywhere within this range of pressures, its height and concentration could be determined. If a cloud were to exist at lower pressures, its precise height could not be determined—it would simply be detected to exist at a pressure <60 mbar. At greater pressures, the inferred aerosol abundance depends primarily on the methane abundance.

Unlike hydrogen, methane condenses in the cold tropospheric temperatures of Uranus' atmosphere; therefore, the mixing ratio of methane potentially varies with height and location. The observed reflectance at wavelengths dominated by methane opacity cannot be used to uniquely constrain the aerosol distribution; however, if the aerosol reflectance is independently determined by using the known hydrogen opacities, and some assumptions are made regarding the wavelength dependence of aerosol scattering properties, then some constraints can be placed on the methane abundance. The total methane abundance above a defined aerosol layer may then be inferred, and the vertical distribution of methane can be resolved if the aerosol abundance at each height is known. This requires observations at two points of the spectrum–one at wavelengths dominated by methane absorption and the other at wavelengths dominated by hydrogen CIA– but with similar transmission functions so that they are sensing reflecting aerosols at the same heights. Deeper in the atmosphere, the hydrogen CIA becomes optically thick and cannot be used to uniquely constrain the aerosol abundance; consequently, the aerosol and methane distribution deeper than a few hundred millibars become ambiguously interdependent. For these pressures, an aerosol profile may be retrieved from our data only if a fixed methane profile is assumed.

Examining Fig. 5, we can see that methane clearly dominates in the H-band and hydrogen CIA dominates in the K-band, but the amount of overlap in transmission functions depends on the methane abundance. If the true atmosphere were similar to the cold-trapped limited case, the methane absorption features at roughly ~1.7 and 2.3 microns would extend high into the hydrogen absorption dominated heights where the aerosol reflectance can be constrained; thus, the methane mixing ratio could be determined between ~150-250 mbar and 70-100 mbar, assuming a relationship between the aerosol scattering properties at the two wavelengths as discussed in Section 3.2.2. For retrieving methane abundances, this would be the best-case scenario. If the methane mole fraction was less, the overlap would be reduced, and the ability to determine the actual value at the same heights would diminish. Inputting the Spitzer value, we find that the transmissions no longer overlap, and so in that situation we could not expect to uniquely resolve the vertical distribution of methane and aerosols nor the precise wavelength dependence of aerosol scattering; instead, the observations would largely be sensitive to the product of the total integrated methane and aerosols above these heights, with only rough vertical discrimination coming from the relative reflectance between the continua and absorptions bands. Through our analysis, we found the Uranus data were more consistent with the latter situation, providing only weak restrictions on the methane profile. Given these limitations, additional modeling constraints on the methane profile were required as discussed in the following sections.

*3.2 Atmospheric Model.*

A 1-D atmospheric model was used to produce synthetic spectra. The model consisted of 70 vertically stacked layers ranging from 20 bars to 1 mbar. Each layer was effectively defined by its methane and aerosol abundance, forming vertical profiles of each variable. Scattering from aerosols were computed assuming plane parallel single scattering as discussed in the following section. Opacities for methane were computed using correlated-k tables with 10 g-ordinates as prescribed by Sromovsky et al. (2012); specifically, these include tables that Sromovsky et al. (2012) produced from their M5 list, derived from HITRAN 2008 and Bailey et al. (2011), with their own far-wing line shape function for the H-band and a Hartmann line shape for a majority of the K-band ($\lambda \geq 2.083$ μm); the remainder of the K-band (2.028 - 2.083 μm) utilized Irwin's tables of correlated-k derived from methane absorption coefficients of Karkoschka and Tomasko (2010), as recommended by Sromovsky et al. (2012) for the inclusion of $CH_3D$ lines. Hydrogen CIA opacities were computed using coefficients of Borysow (1996), adapted for arbitrary para-hydrogen fractions; the fraction was initially a free parameter, but subsequent testing demonstrated that the precise values could not be constrained by this data set, and it was thus set to the local thermodynamic equilibrium fraction. Since the gaseous absorption is a function of pressure and temperature, a standard temperature-pressure profile from Voyager radio occultation data (Lindal et al., 1987, model F) was used with a constant helium mole fraction of 0.152 (Conrath et al., 1987); other published profiles were tested (e.g. Lindal et al., 1987 model D; Sromovsky et al., 2011), but the results were not significantly sensitive to differences in these temperature profiles. For simplicity and computational efficiency, the variable methane mole fraction was neglected from calculations of each layer's mean molecular weight and hydrogen abundance, such that the hydrogen mole fraction was not decreased when the methane mole fraction was increased. Testing showed that this approximation produced no significant difference in results when compared to cases in which the hydrogen fraction was accordingly reduced.

### 3.2.1. Aerosols Scattering and Reflectance

Aerosol distributions were modeled following methods of Banfield et al. (1996, 1998a) and Stam et al (2001). Each layer of the atmosphere had an optical thickness of gas and aerosols; the aerosol component scattered light and resulted in the observed reflectance assuming a single-scattering regime. Light scattered from a single aerosol is assumed to have entered the atmosphere at the local solar incidence angle and emitted at the local emission angle towards the observer.

In a single-scattering regime, the contribution to the observed reflectance from each layer is dependent on the number of aerosol particles in the layer, $n$, the effective cross sectional area of each particle, $\sigma$, the single scattering albedo of the aerosols, $\varpi_0$, and the normalized scattering phase function evaluated at the angle between light incidence and emission, $PF(\alpha)$. The product of these terms is here referred to as the scattering optical thickness, $\tau_s$ :

$$Eq.\,1) \qquad \tau_s(z, \lambda) = n(z)\, \sigma(z, \lambda)\, \varpi_0(z, \lambda)\, PF(\alpha, z)$$

Where $z$ and $\lambda$ designate that parameters are a function of height and wavelength. The total reflectance due to single scattering can be found by vertically integrating over all layers down to sufficiently great atmospheric optical depths:

$$Eq\,2.) \qquad I(\lambda) = \frac{F(\lambda)}{4} \int_0^\infty n(z)\, \sigma(z, \lambda)\, \varpi_0(z, \lambda)\, PF(\alpha, z)\, e^{-\tau(z,\lambda)\mu_0}\, e^{-\tau(z,\lambda)\mu}\, \frac{dz}{\mu}$$

Where $I(\lambda)$ and $\pi F(\lambda)$ are the observed and incident solar flux at wavelength $\lambda$, $\tau(z,\lambda)$ is the total atmospheric optical depth from the top of the atmosphere down to height $z$, and $\mu_0$ and $\mu$ are the cosines of the incidence and emission angles, respectively. Assuming hydrostatic balance, the height could be expressed in pressure coordinates. Given a range of atmospheric weighting functions, one can potentially invert the observations of $I/F(\lambda,\mu,\mu_0)$, to retrieve the scattering optical thickness, $\tau_s$, as a function of pressure for each wavelength.

The scattering properties of Uranus' aerosols are likely wavelength dependent (e.g. see Irwin et al, 2007, 2009, 2010, 2011, 2012; Karkoschka and Tomasko, 2009; Sromovsky et al, 2011; Tice et al, 2013). The effective scattering cross section, single scattering albedo, and scattering phase function could all potentially contribute wavelength dependence to $\tau_s$; however, any potential wavelength dependence would only be very weakly constrained by our observations since aerosols at any given height are only dominating the reflectance over small subsets of the entire spectrum. The deeper cloud layers contribute the greatest reflectance in H band, but only over a relatively small window around 1.6 µm, hence there is little information on how reflectance varies with wavelength. The K-band and strong absorption features of the H-band offer sensitivity to heights near the tropopause over a broader range of wavelengths, but the data are simply too noisy to place meaningful constraints on all three scattering parameters.

Rather than allowing each of the three under-constrained scattering parameters to vary freely, we attempted to model any necessary wavelength dependence in aerosol opacity using a single parameter– specifically, a trend in aerosol extinction efficiency versus wavelength, corresponding to a particular particle size. Hypothetical trends in aerosol extinction efficiency were computed using Mie theory for 100 different particle sizes ranging from 0.05 to 5 µm in radius. These trends were normalized at 1.48 µm, meaning only the ratios of aerosol opacities at longer wavelengths relative to 1.48 μm could change. Retrievals were systematically performed using each of the potential particle sizes as a proxy for wavelength dependences, and Chi-squares were evaluated and compared.

Both single scattering and multiple scattering codes were tested for computing the model reflectance. In theory, single scattering calculations are sufficiently accurate for modeling reflectance from clouds and hazes with a total optical thickness less than ~0.1 (van de Hulst, 1957). In a cloud of greater optical thickness, the projected area of individual particles begins to overlap, and the resulting reflectance is not simply proportional to the number of particles. In these circumstances multiple scattering code is necessary for calculating the correct reflectance. Preliminary retrievals show that H-band scattering optical depths begins to exceed 1/10 at about 1.8 bars, suggesting multiple scattering occurs below this height. At these depths, the methane abundance is greater, and the increased path length of multiply scattered photons may potentially mitigate the contribution of multiply scattered light if the distances between scatterings are long enough; however, a simple test shows that doubling the scattering optical thickness of this layer produces a less-than twofold increase in reflectance, thus suggesting multiple scattering is indeed significant for the deeply penetrating regions of the H-band. To test the sensitivity, results from representative examples were retrieved using both single and multiple scattering calculations, using a code based on algorithms of Hansen and Travis (1974) and used previously by Banfield et al. (1998b) and Roman et al. (2013). Multiple scattering calculations require that individual scattering properties—bundled together in the above $\tau_s$—be specified, since these properties are necessary for evaluating the contribution from subsequent scatterings. Figure 6 shows one example demonstrating that the retrieved aerosol scattering optical thickness for a deep cloud layer can be up to 45% less when using multiple scattering calculations in the case of conservative, isotropic scattering (in this test case, the multiple scattering code used a $\varpi_0 = 1.0$, a Henyey-Greenstein phase function with asymmetry parameter of 0.1, and a spherical harmonics method with 20 azimuth divisions, 4 Gauss divisions, and 3 Fourier terms, as described in Hansen and Travis (1974)). Different scattering assumptions produce somewhat different results. Non-conservative scattering and anisotropic scattering tend to reduce the reflectance of the clouds in both cases, but to a greater extent in the multiple-scattering case (where each scattering leads to additional absorption); consequently, the differences between multiple and single scattering results depend on the scattering properties, and the conservative, isotropic case serves as a limiting case. So while multiple scattering calculations are more accurate, uncertainties in the necessary scattering parameters still result in significant uncertainties in the retrieved magnitude. For the present investigation, the value of knowing the aerosol distribution rests in the qualitative picture and relative changes as opposed to absolute values of retrieved parameters. Since the multiple scattering calculations are computationally very costly, the added computational time was deemed too burdensome given the uncertainties. Hence, single scattering was assumed, and reported deep cloud scattering optical thicknesses may be interpreted based on preferred scattering assumptions.

Our retrieved scattering optical depths as reported throughout this paper should be interpreted as the product of the scattering terms (Eq. 1). For example, as reported without conversion, our stated $\tau_s$ bar$^{-1}$ would represent the optical depth (i.e. number density x extinction cross-section, per bar) for a cloud of particles with a single scattering albedo of and normalized phase function of 1.0. For a cloud located at several few bars, reasonable values based on the literature (Tice et al., 2013; Irwin et al., 2015), as discussed in Section 5.2, would suggest single scattering albedo of ~0.7 and a normalized phase function of ~ 0.4, yielding an optical depth ~3x our reported $\tau_s$ bar$^{-1}$.

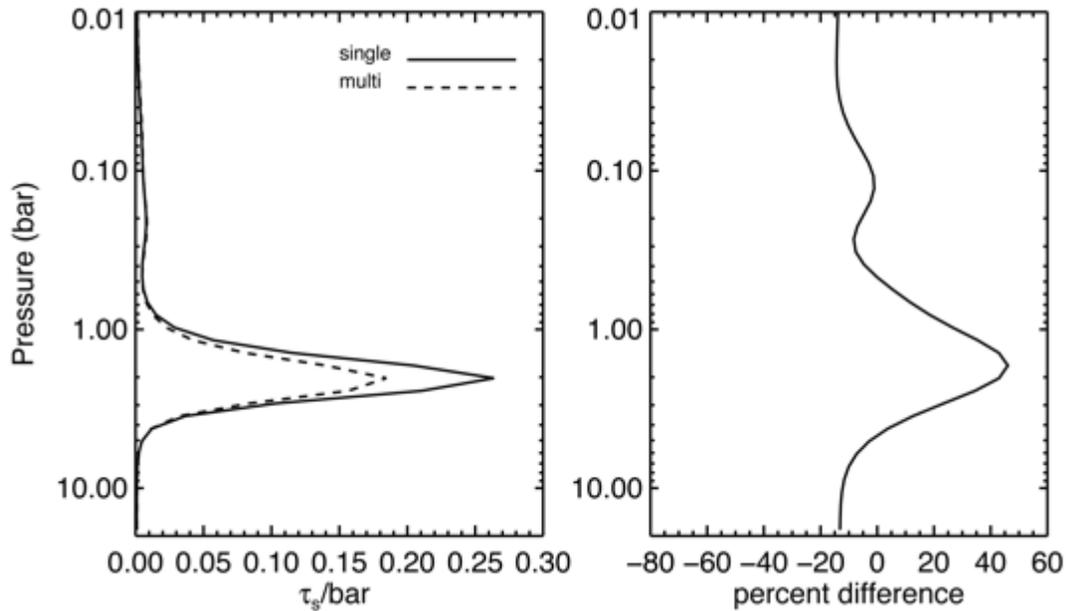

Figure 6. Comparison of retrievals from real data performed using both single and multiple scattering. Calculations assumed the same conservative, isotropic scattering phase function, serving as a limiting case. The plot on the left shows that less scattering optical thickness is required to produce the same reflectance; the difference is most pronounced in the optically thickest cloud. The plot on the right shows the difference as a percentage of the multiple scattering results.

*3.2.2. Methane profile.*

To account for uncertainties in the methane vertical profile, several plausible methane models were separately used to retrieve the resulting aerosol profiles. For pressures greater than ~1 bar, different deep methane mole fractions and latitude dependences were evaluated. For pressures less than ~1 bar, aerosols were retrieved for two different profiles with fixed mole fractions and one profile that treated the stratospheric mole fraction as free parameter.

Published values for the deep methane mole fraction include 2.3% from Lindal et al. (1987), 3.2% from Karkoschka and Tomasko (2009, 2011), and 4% from Sromovsky et al. (2011) and Sromovsky et al. (2014). The latter studies also suggested equator-to-pole variation, with methane depletion at the pole. To evaluate the data's sensitivity to differences in the deep methane models, aerosol profiles were retrieved and compared for three different assumptions. The first assumed a constant mixing ratio of 4% at pressures greater than 1.4 bar for all latitudes (Sromovsky et al. 2011); the second case was latitude dependent, following Karkoschka and Tomasko (2011) with a fixed deep abundance of 3.2%, but with the height of this layer pushed deeper towards the poles (from their Figure 10). The third case was based on Sromovsky et al (2014), who argued that the polar depletion could not extend to great depths, and assumed a 4% deep abundance with a different latitude dependent height (see their figure 17). In all assumptions, a north-south symmetry was assumed. It should be note that such an assumption is an oversimplification, since studies have found a north-south asymmetry in the methane profile that may affect the retrieved aerosols optical depths (Irwin et al., 2012; Sromovsky et al., 2014; de Kleer et al., 2015). This resulted in a family of profiles like those illustrated in figure 7.

To complete the profiles, the deep methane value was smoothly interpolated in logarithmic pressure from ~500 mbar (where the different models roughly converged) up to a 200 mbar; from the 200 mbar up to the top of the model was modeled for three different configurations. In the first setup, the methane mole fraction was fixed at $1.6 \times 10^{-5}$ from 200-mbar to the top of the model, based on Spitzer results (Orton et al., 2014). In the second setup, the tropopause mole fraction decreased from $\sim 8 \times 10^{-5}$ at 200 mbar to the Spitzer value at 2 mbar, following the Herschel/PACS results (Lellouch et al., 2015, their figure 1). In the third setup, a uniform mole fraction from 200 mbar to the top of the model was left as a free parameter to be fitted by the data. Plots of these methane profiles on a logarithmic scale can be seen in Fig 12. In all cases, the mole fraction was prevented from exceeding saturation, as determined by the saturation vapor pressure and temperature profile 'F' (Lindal et al, 1987).

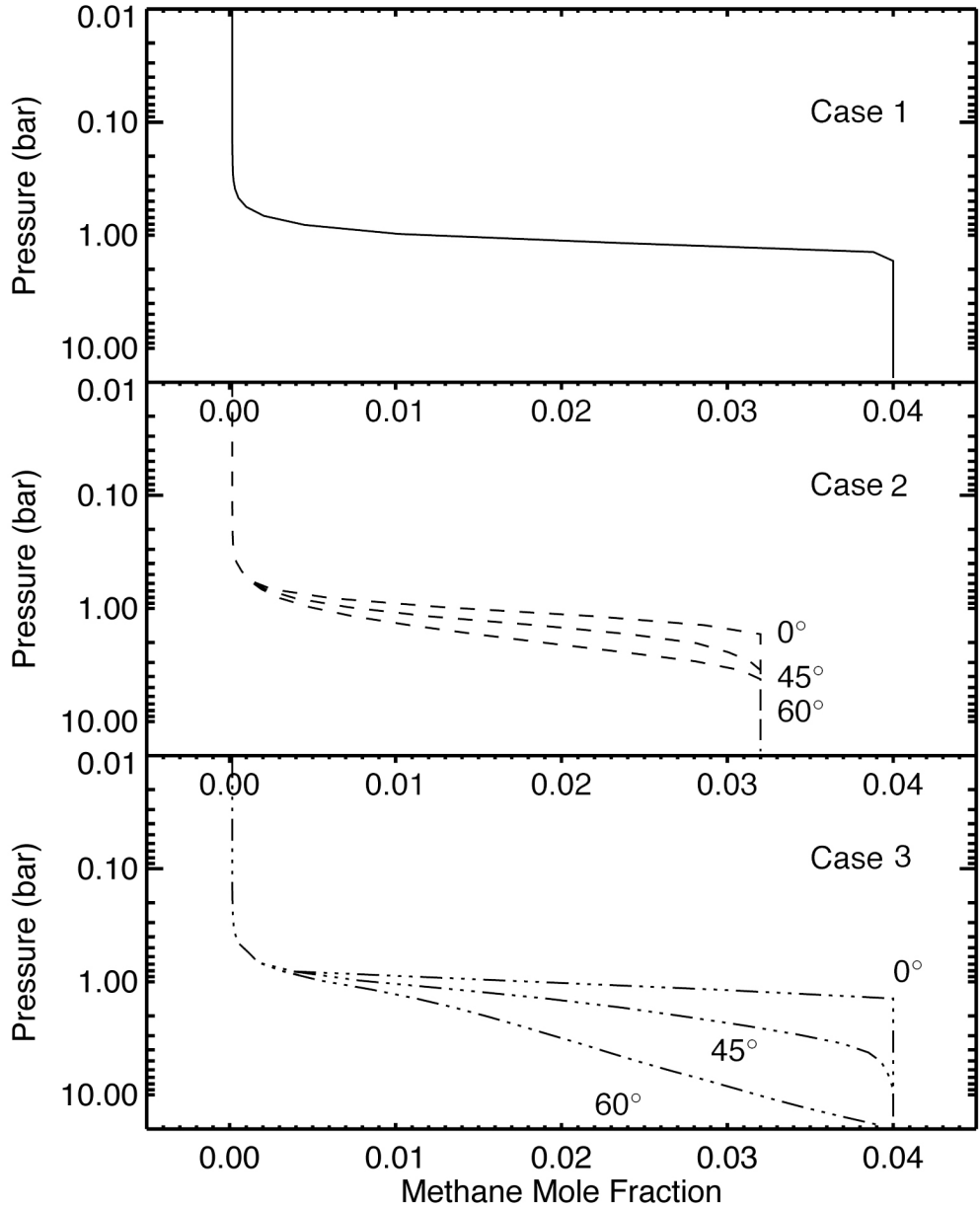

Figure 7. Three different models for the deep methane distribution to be evaluated by the retrievals. Cases 2 and 3 assume latitude dependence, while Case 1 does not, as discussed in the text. Methane profiles at 0°, 45°, and 60° latitude are illustrated.

Note that no effort was made to ensure that methane profiles were self-consistent with retrieved cloud condensation heights or the inferred cloud height in the Voyager radio occultation results. Indeed, given the Lindal et al. (1987) temperature profile, none of these methane profiles reach saturation on average. This apparent incompatibility may be resolved by reducing the temperature profile, as shown possible by the Sromovsky et al. (2011) reanalysis of the Voyager refractivity data, or by simply regarding the profiles as spatial averages of saturated and unsaturated regions, as Lindal et al (1987) suggested.

*3.3 Retrieval Algorithm*

Constrained vertical profiles of aerosol scattering optical thickness per bar of pressure were retrieved from both H and K spectra using an inversion algorithm, following Banfield et al (1996,1998a) and Stam et al (2001). For most cases, the methane profile was fixed as discussed in the previous section. For cases when the stratospheric methane mole fraction was treated as a free parameter, the mole fraction and aerosol profile were retrieved simultaneously. The method aimed to retrieve the simplest solutions possible while minimizing differences between observed and modeled spectra.

The modeled spectra matched the overall observed spectra well within errors, though there were some systematic errors (Fig 8). The model spectra were consistently 10-20% too dark in regions of the H band between ~1.53-156 μm. These errors were beyond the observational uncertainties. This may be due to errors in the methane absorption coefficients or possibly a complication in vertical structure that exceeded our model's complexity. The K band spectra were very noisy with large errors bars that accommodated a range in modeled spectra. The error bars at these wavelengths were proportionately large and thus did not have significant influence in the error-weighted fitting algorithm.

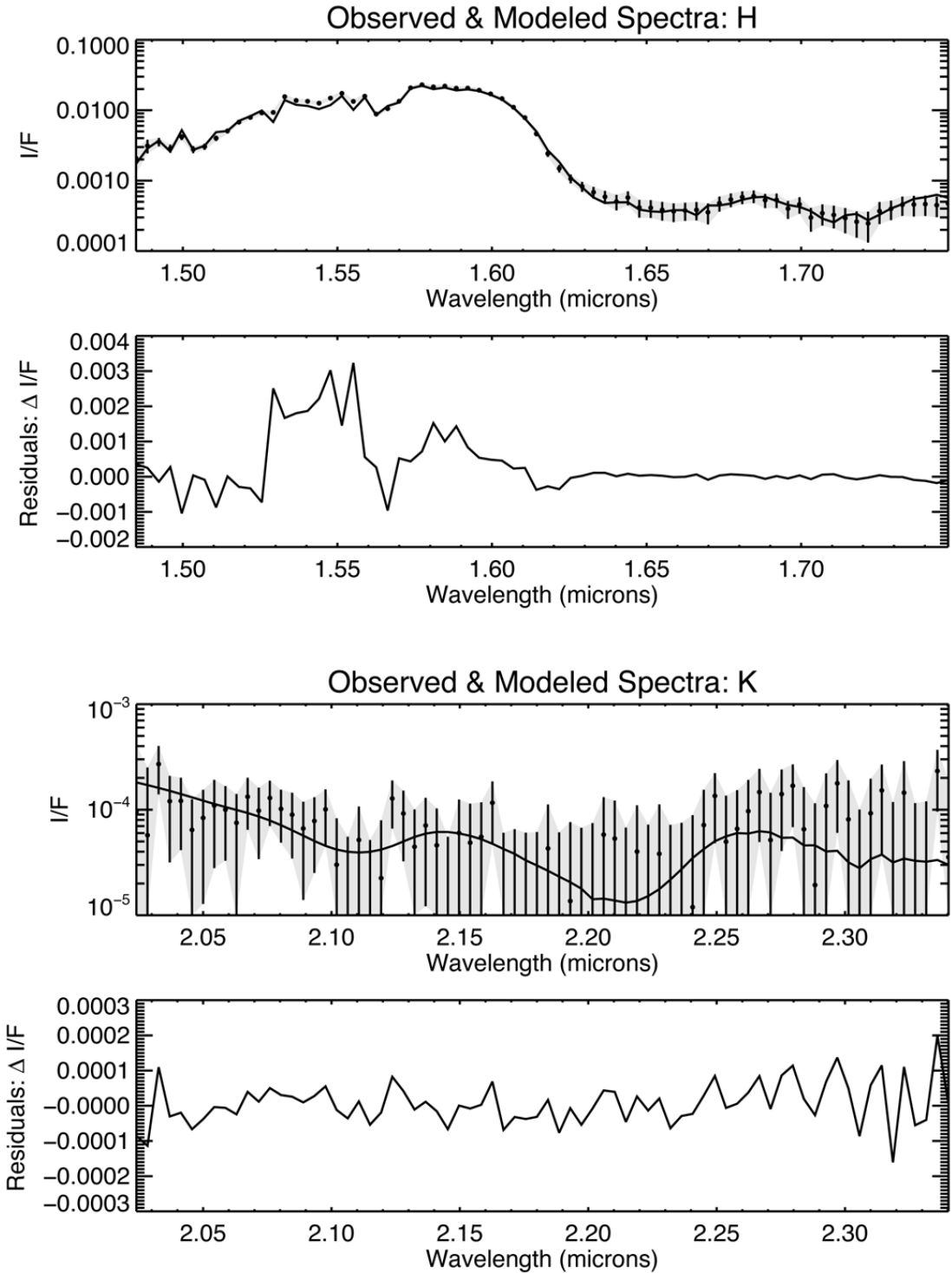

Figure 8. Observed and modeled spectra for a best fit, along with residuals between data and model, for H (top) and K (bottom) bands. Irradiance is expressed in unit-less I/F. Dots represent observed values, vertical lines and accompanying gray shading express the uncertainties, and the solid lines are the model spectra and residuals (observation – model).

Small perturbations to the free parameters at each height were used to compute the sensitivity of the modeled spectrum at each wavelength. Based on these linearized derivatives, an improved profile was computed through minimization of a modified quadratic merit function. Unlike the typical merit function that is minimized in least squares fitting, this modified function contained additional terms that constrained the solutions to lay close to reference values—in this case, the previously estimated values. Numerically, changes to the profile carried a penalty that increased the merit function while improvements to the fit reduced the merit function; the importance of these opposing terms were determined by carefully tuned weighting factors. The additional terms served to filter the solutions, preventing arbitrarily small changes in the modeled spectrum from mapping into spurious large changes in weakly constrained parameters. A correlation length scale of half a scale height was imposed between adjacent layers to further restrain the solutions. The profiles were evolved iteratively, as new functional derivatives were computed each iteration to account for non-linearity in the solutions. Fits were stopped after 15 iterations, as iterating further did not improve the fit significantly beyond the noise. More details on the retrieval algorithm formulation can be found in Section 2 of Conrath et al (1998).

Weights for the smoothness constraints were carefully optimized based on tests of synthetic data and the information content of the real data. Our goal was to find the smoothest solutions that adequately match the observed spectrum to within random measurement error. Starting with weak constraints, the weighting values were gradually increased to dampen any unnecessary ripples in the profile that would imply needless complexities in vertical structure. The values of the optimal dimensionless weighting factors were weakly dependent on the vertical structures and resulting intensities of the spectra. For a majority of the data, we found a consistent weighting factor provided optimal constraints, balancing the goodness of the fit with the complexity of the vertical profiles. Factors of two or three times greater were required for some perversely devised test cases.

*3.3.1 Retrieval Characteristics and Uncertainties.*

The general accuracy and nature of the retrieved solutions can be seen in attempted retrievals from synthetic data. Hypothetical profiles were devised and used to forward model spectra, to which realistic noise was added. Profiles were then retrieved from the synthetic spectra and compared to the actual source profiles.

As can be seen in fig. 9, the retrieved solutions of aerosol profiles do not represent a literal distribution of aerosols, but rather represent a modeling construct that captures the basic structure—specifically the peak heights and scattering optical thicknesses of aerosol layers. The algorithm yields solutions of forms that are smooth and rounded in nature and intrinsically limited in retrieved vertical resolution by the data. Discrete cloud layers are retrieved as rounded features vertically centered on the cloud height and roughly conserving the total optical thickness in best cases. The pressures and magnitude of the peak scattering optical thicknesses are captured well, particular for a deep, optically thicker layer, with typical errors in height and $\tau_s$ bar$^{-1}$ of 10% or less.

Extended high aerosol layers of low optical density are not fit nearly as well, particularly if they extend well above 100 mbar. Towards the top and bottom of the model, the profiles converge to the initial values as the functional derivatives becoming vanishingly small. If hazes extend into these rarified heights, the retrieval will truncate the top as it converges back to the initial profile. The peak $\tau_s$ bar$^{-1}$ of a high haze is usually reproduced to within 30% for pressures greater than ~50 mbar, but could be off by an order of magnitude for devised cases of very thin, discrete haze layers, with very low integrated optical thickness; however, for models actually consistent with the observed spectra (i.e. solutions to the real data), the forward modeled spectra can be inverted to retrieve the upper haze height and thickness to within an accuracy of 20%.

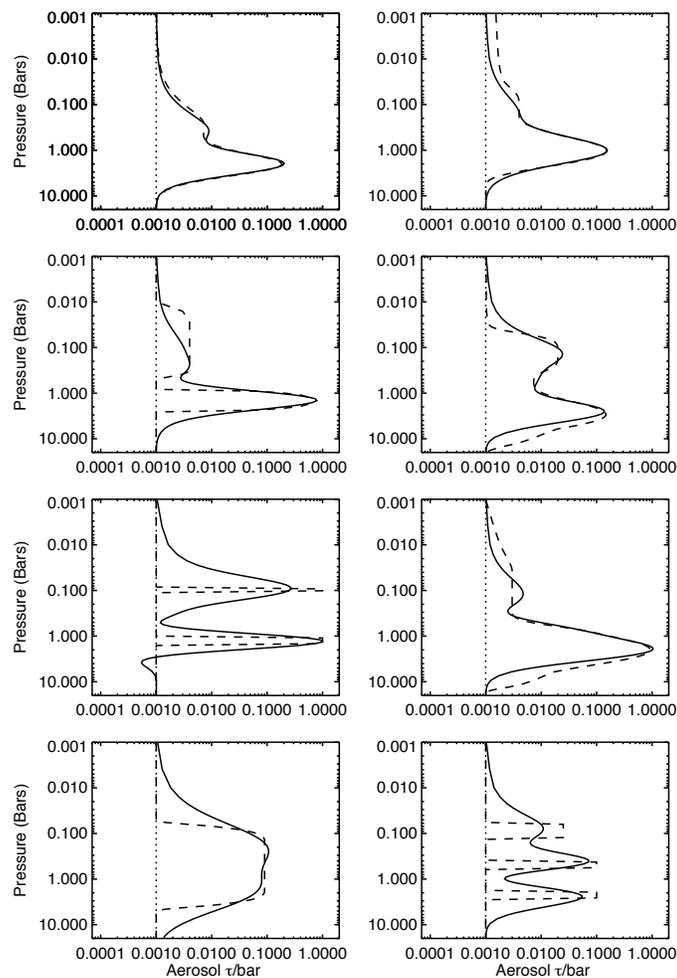

Figure 9. Eight representative examples of retrievals of aerosols from synthetic data. Each dashed profile is a devised distribution that was forward modeled to produce synthetic spectra, to which realistic noise was added. The spectra were inverted using the retrieval algorithm to yield solution for the vertical aerosol distribution (solid line), initiated at the dotted line.

The remainder of each retrieved aerosol profile does not necessarily represent the reality and should not be considered significant aside from its contribution to the total integrated reflectance. The precise shape is dictated more by the retrieval algorithm than by the data. Aside from the upper and lower margins, we find little sensitivity to initial aerosol profile so long as it is does not far exceed the final profile in reflectance.

Attempted retrievals of the stratospheric methane mole fraction are less robust. Retrieved values can differ from the actual source value by a factor of three, particularly when either the true methane value or aerosol optical thickness is very low. This is likely due to the very low signal to noise of the K-band observations and the inability of the data to break degeneracies when the weighting functions in methane bands peak below the hydrogen weighting functions. Indeed, we often found a correlation between the retrieved stratospheric aerosol optical thickness and methane mole fractions, aggravated further with increasing noise. In such cases, the retrieved methane values were weakly sensitive to initial conditions.

For aerosol models consistent with the observed spectra, the retrieved methane mole fractions were consistently accurate to within a factor of two or less. For models with greater aerosol concentrations (akin to locations with discrete clouds), the retrievals performed consistently better; however, we still observed a slight correlation between the methane mole fraction and the aerosol optical thickness, suggesting persistent degeneracy in the solutions. Consequently, if larger methane mole fractions are assumed and held constant, the resulting aerosol profiles can compensate by increasing the scattering optical thickness with only modest increases to the Chi-squared.

To further evaluate the sensitivity of the spectrum to the upper methane profile, we compared the goodness of fits to data for different stratospheric mole fractions retrieved in conjunction with aerosol profiles (see fig. 10). For a given observation, we found that 200-mbar mole fractions ranging from $1.2 \times 10^{-5}$ to $2 \times 10^{-5}$ yielded essentially equivalent Chi-squares. The profile above these heights had no significant effect on the Chi-squares. From these tests, the retrieved methane mole fraction between 100 and 200 mbar should only considered to be accurate to within a factor of 2, with error bars of no less than $2 \times 10^{-5}$. Well above these heights, the data does not constrain the value

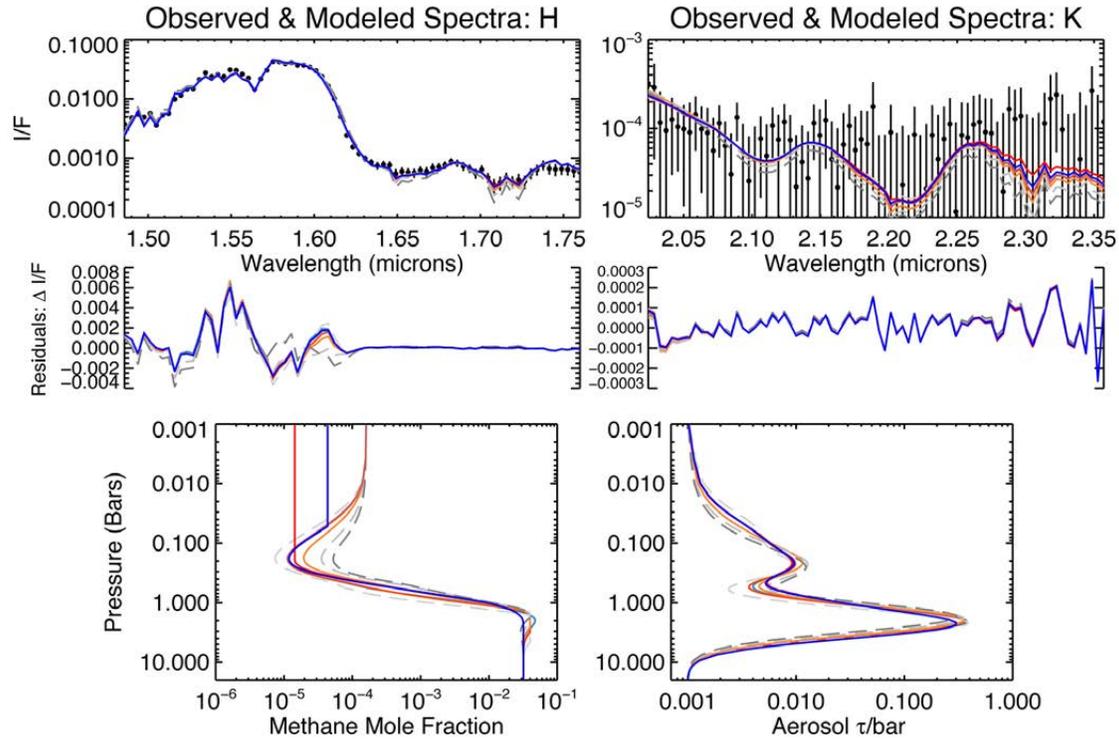

Figure 10. (top) Example of observed and modeled spectra (26˚S latitude on September 3, 2001). The data are represented by black dots, the vertical lines represent the error bars, and the solid lines represent the various good fits with comparable residuals made using different assumptions and constraints on the methane profiles; the dashed lines represent similar but poorer fits. (bottom) The corresponding vertical structures of methane mole fraction and aerosol scattering optical thickness per bar for each of the above spectra, in corresponding shades and line styles. The spread of acceptable models illustrates the intrinsic sensitivity of the data and provides an additional estimate of uncertainty in the retrievals.

Though the methane profile was overall weakly constrained and largely degenerate with solutions to the aerosol profile, the goodness of the fits still depended, albeit slightly, on the choice of methane models, suggesting the data can provide some slight discrimination between parameter choices within our assumed modeling framework. These conclusions are presented along with the results.

*4. Results*

Using the above methods, we retrieved scattering optical thicknesses per bar as a function of pressure and latitude for each observation in the data set. Retrievals were completed using several different methane models; for one set of solutions, we attempted to simultaneously retrieve a uniform methane mole fraction as a function of latitude for pressures less than 200 mbar.

Statistical Chi-squares of modeled fits to the observations were minimized, and minimum values for different cases were compared to evaluate different methane models. The Chi-squares can be reduced by the degrees of freedom to yield the reduced Chi-squares, but we note that the degrees of freedom are not well defined for non-linear models (Andrae et al 2010). If we estimate the degrees of freedom by the number of binned data points (~150) minus two parameters (one for each retrieved profile), the reduced Chi-squares range in values from approximately 0.5 to 1.2 on average for our best fitting models. Typically, values less than 1.0 suggest that the uncertainties or degrees of freedom are slightly overestimated, but we simply use the Chi-squares as a consistent measure of the relative magnitude of residuals for different cases.

*4.1 Preferred Methane Models*

We compared values of Chi-square using each of the different methane models discussed in Sec. 3.2.2. The deep methane models with latitude dependence provided much better fits at high latitudes than did the uniform deep profile. This can easily be seen in Fig. 11, which illustrates the Chi-squares as a function of latitude for each of the deep methane models given the same stratospheric / upper tropospheric methane model. The Chi-squares of the latitude dependent models of Karkoschka and Tomasko (2011) and Sromovsky et al (2014) deviate from the latitude independent values at ~30° N and S throughout all the data. This clearly indicates that the data have some constraining sensitivity to the deep methane model between ~1 and ~3 bars even when the aerosols profiles are largely free to adjust, and supports previous observations of methane depletion towards the poles at pressures of a few bars. The differences in Chi-squares between the two pole depleted models are insignificant, with the model of Karkoschka and Tomasko (2011; case three in our Fig. 7) providing insignificantly lower Chi-squares on average. Accordingly, throughout the remainder of this paper all presented results were calculated using the Karkoschka and Tomasko (2011) deep methane model.

For heights above the 1-bar pressure level, the three upper methane models discussed in Sec. 3.2.2 were evaluated. Figure 12 shows these models, including the average retrieved results for each year analyzed. As with the deep models, Chi-squares were computed for each case as a function of latitude, while using the same deep methane model (i.e. the latitudinal dependent model of Karkoschka and Tomasko, 2011) and allowing the aerosols to freely adjust; Chi-squares for each case and latitude are plotted in Fig 13. In that figure, the solid line represents the setup in which the uniform mole fraction above 200 mbar was treated as a free parameter (initiated at the Spitzer value of 1.6 x $10^{-5}$). The dashed line represents the fixed profile based on Spitzer results (Orton et al., 2014), and the dot-dashed line represents the fixed, more humid, profile from the Herschel/PACS results (Lellouch et al., 2015). Retrievals of aerosol profiles using the Spitzer and free-parameter models were consistently slightly better—approximately 9% on average—than those computed using the Herschel/PACS profile. The Chi-squares of the free-parameter fits do not differ appreciably from those for the fixed Spitzer model. Indeed the retrieved mole fractions are roughly consistent to the Spitzer value of (1.6 +.2 /-.1) x $10^{-5}$. The mean retrieved stratospheric mole fraction was 1.2 x $10^{-5}$ with a standard deviation of 0.3 x $10^{-5}$, though as discussed above, the actual uncertainty may be as large as a factor of two given the potential for degeneracies between the methane and aerosol abundances.

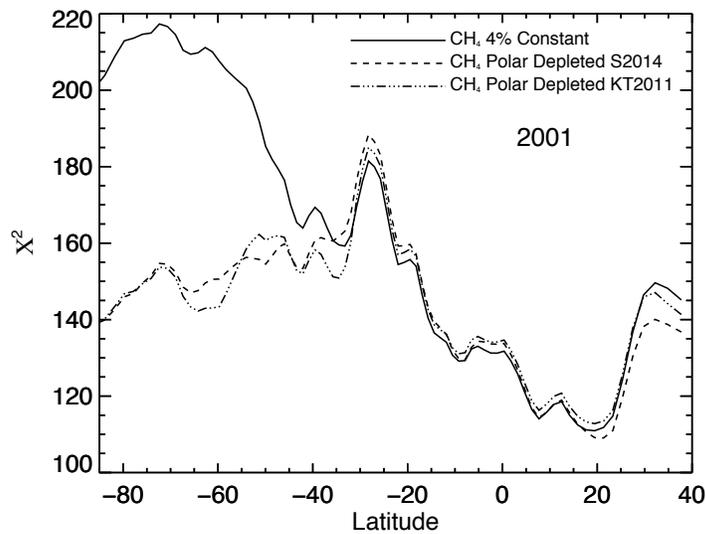

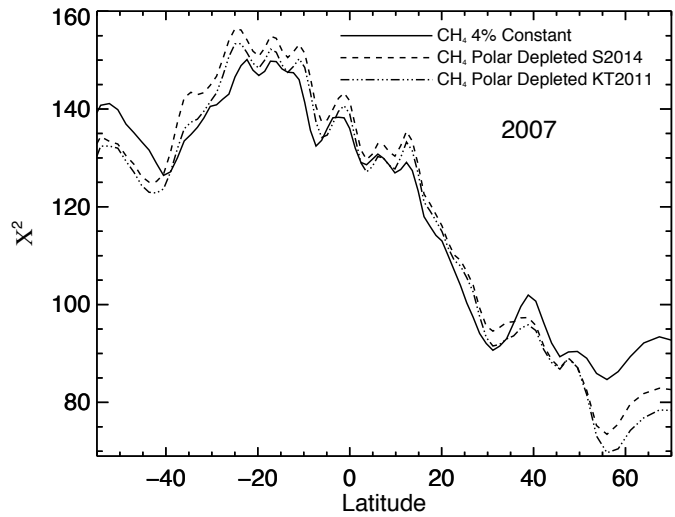

Figure 11. The average goodness of the fits, expressed as Chi-squared, as a function of latitude for three different deep methane models and two different years–2001 and 2007. In both cases, the latitude dependent models (dashed and dot-dashed lines) provide better fits than the latitude independent model (solid line) for latitudes pole-ward of ~30˚. The Chi-squares were computed from spectra, each binned into 154 data points in 2001 and 146 in 2007.

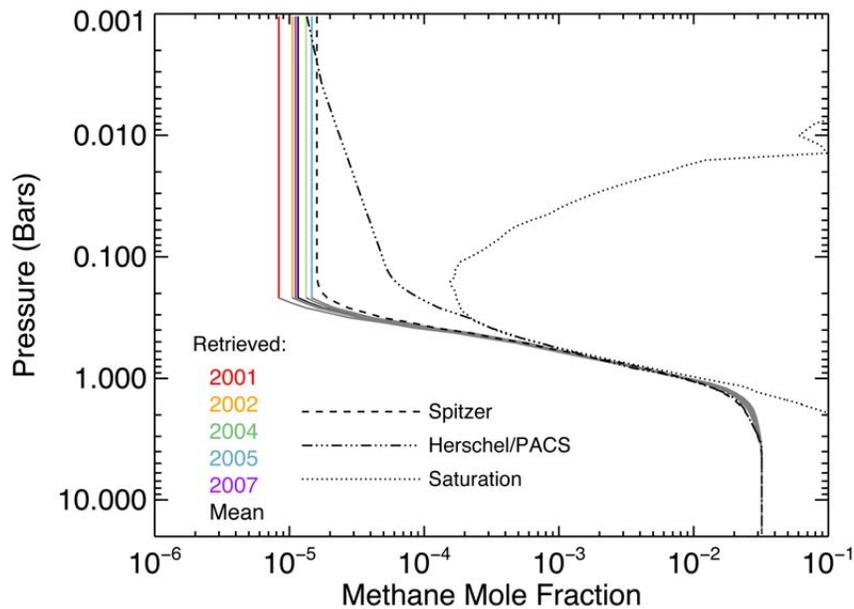

Figure 12. Methane mole fraction as a function of pressure for different methane models. The dot-dashed line represents the inferred profile from Spitzer results (Orton et al., 2014). The dashed line is the profile suggested by Herschel/PACS. The solid lines

represent fits to the data using a profile with a free parameter, with values color coded indicate the year and the mean value in black. The difference between years is less than the measurement uncertainty. The dotted line indicates a saturated mole fraction based on the temperature profile and saturation vapor pressure.

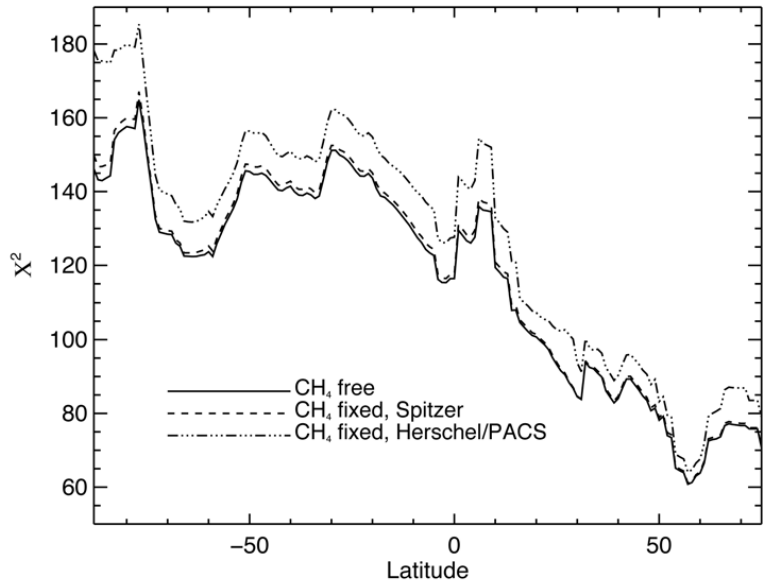

Figure 13. The goodness of the fits for three upper methane models, expressed as Chi-squares, as a function of latitude averaged over all the observations. The solid line represents the free-parameter uniform mole fraction above 200 mbar. The dashed line represents the fixed profile based on Spitzer results (Orton et al., 2014), and the dot-dashed line represents the profile from the Herschel/PACS results (Lellouch et al., 2015). Retrievals assume a deep methane model of Karkoschka and Tomasko (2011). On average, each spectrum was binned into ~150 data points, from which we estimate reduced Chi-squares ranging from approximately 0.5 to 1.2.

A comparison of the fits using the three different upper-methane models is shown in fig. 14. In this example, the differences between the fits are small, with the Chi-square being only 6% greater for the Herschel/PACS profile fits compared to the free-parameter case. Yet the corresponding retrieved profiles differ significantly. The Herschel/PACS methane mole fraction is roughly 7x greater at ~200 mbar than the retrieved mole fraction, but this is evidently largely compensated by a corresponding aerosol scattering optical thickness that is more than 50% greater.

These differences lead to discrepancy in the spectra at wavelengths with very low signal to noise, and hence have minimal affect on the error-weighted Chi-square. Though the Spitzer and free-parameter solutions are marginally preferred, the small difference in Chi-squares suggests that the retrievals are still tainted by slight degeneracies between the aerosol and methane abundances at these heights; consequently the data cannot strongly discriminate between the different methane models. Accordingly, results assuming each of the three upper methane models are discussed.

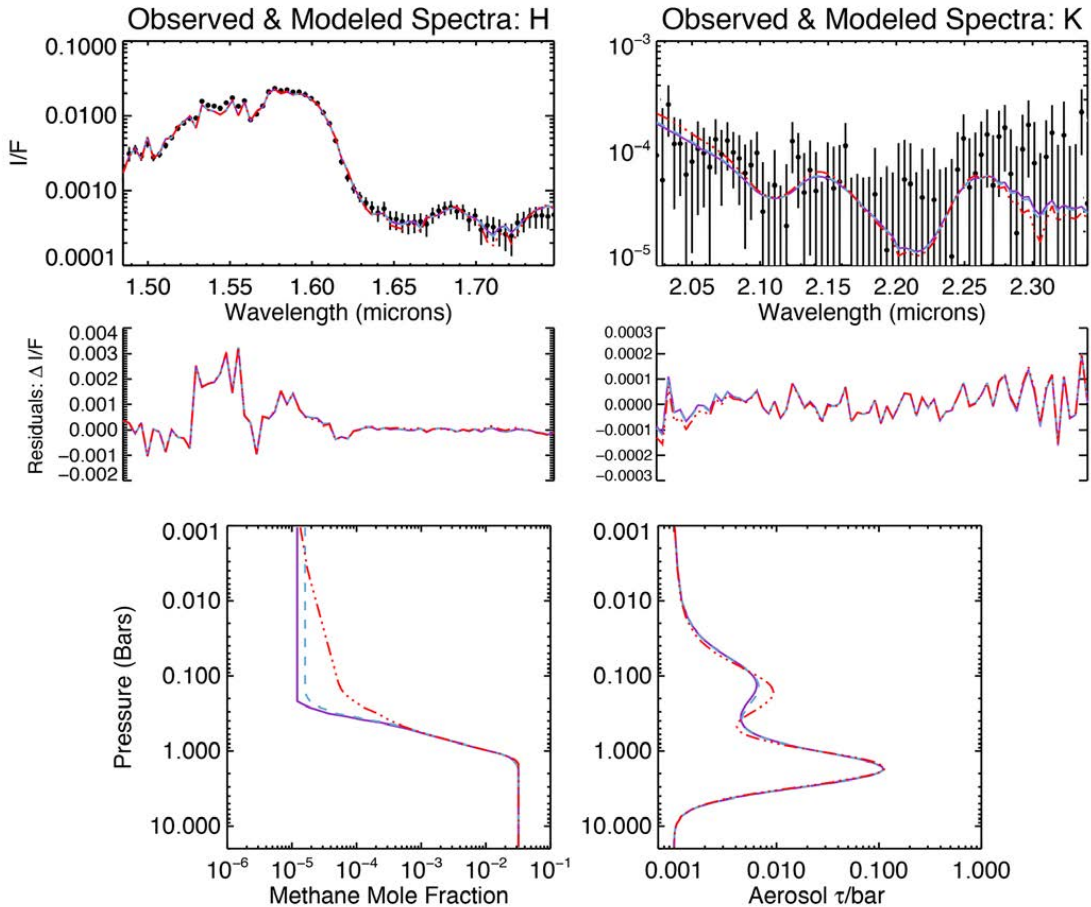

Figure 14. (top) Observed and modeled spectra, along with residuals, for three retrievals from the same data using the three different upper methane profiles. Each modeled spectrum is color coded to the corresponding methane and retrieved aerosol profiles

(bottom). Despite using three different models, the fits are similar, with only slightly larger residuals in the strongest methane features of the H band and throughout the noisy K band. The retrieved aerosol profiles differ significantly between 100 and 300 mbar and yet both provide a reasonable fit.

*4.2 Mean Retrieved Aerosol Profiles*

Vertical profiles of the scattering optical thickness per bar of pressure (as defined in Sec. 3.2.1 and expressed at 1.48 µm) were retrieved for latitudes captured in the slit in each observation. Locations near the edge of the disk were omitted to reduce the errors associated with imperfect atmospheric seeing (as discussed in Sec. 2.2). Average results using each of the three upper methane models are shown in fig. 15.

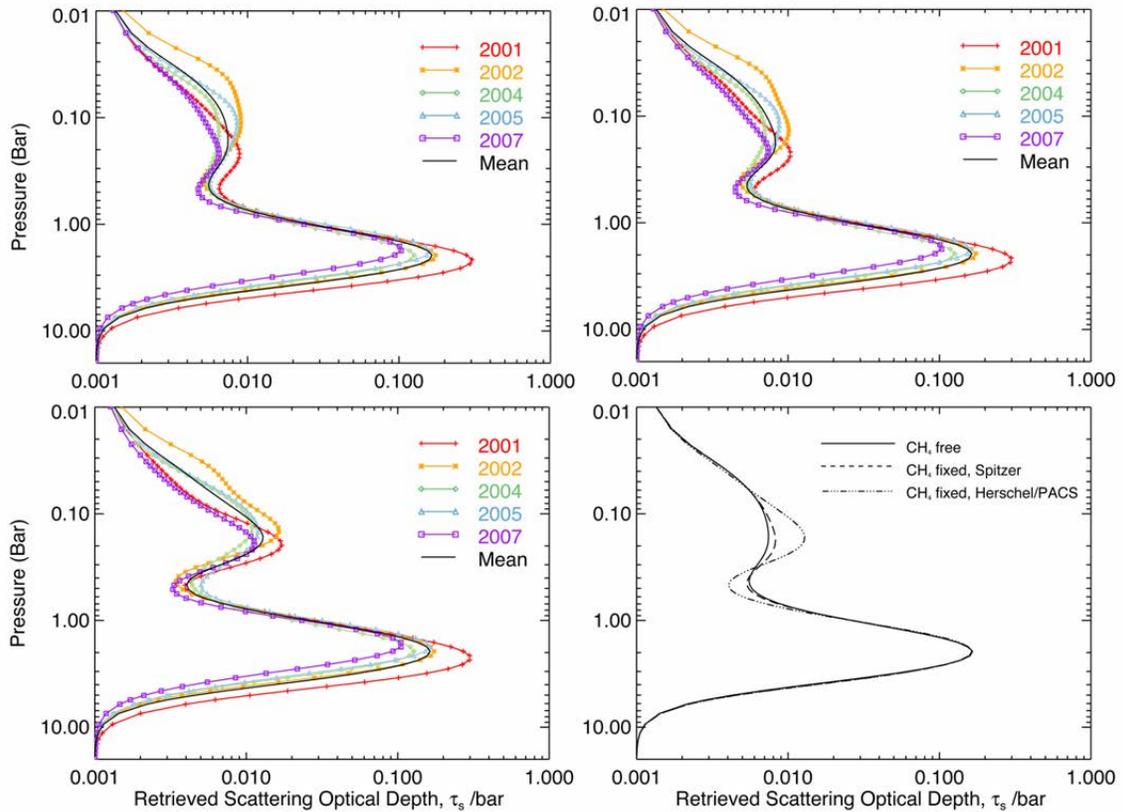

Figure 15. Mean retrieved aerosol profiles for each year a) with the methane mole fraction above 200 mbar as a free parameter, b) fixed Sptizer profile, and c) fixed Herschel/PACS profile. d) Average retrieved profile for each model.

Over time and latitude, the vertical profiles of aerosols consistently displayed two distinct layers: an very optically-thin haze near the tropopause and a deeper, thicker haze or cloud layer as found in other NIR spectral studies (e.g. Irwin et al. 2010; Tice et al., 2013). The deeper layer had an average peak scattering optical thickness per bar of 0.16 at 1.94 bars with a total integrated scattering optical depth of 0.25 between ~0.5 and ~3 bars, beyond which the atmosphere becomes opaque.

Assuming a methane mole fraction of $1.6 \times 10^{-5}$ at heights above 200 mbar, the retrieved aerosol abundance for the top layer had a broad peak centered at 160 mbar with a scattering optical thickness of 0.007 per bar. The integrated optical thickness down to 500 mbar was only 0.003 on average. These values are not significantly different from the mean results found using a free parameter for the methane mole fraction above 200 mbar (that mole fraction was found to be $1.2 \times 10^{-5}$, accurate to within a factor of two, as reported in the previous section); for that case, the layer had a similar integrated optical thickness, peaking at 180 mbar with an peak optical thickness of 0.008 per bar. Finally, if the Herschel/PACS profile suggested by Lellouch et al. (2015) is applied, the increased methane mole fraction yields a peak optical thickness per bar of 0.013 at 160 mbar, with a ~15% greater integrated optical depth.

As expected, the sensitivity to the aerosol wavelength-dependent extinction efficiency was very limited. The data were marginally best fit using a relatively flat dependence with only slight decrease towards longer wavelengths. Assuming Mie theory applies, this corresponded to scattering aerosols with a mean effective radius of roughly one micron or greater at pressures of a few hundred millibars. Particles much smaller than this produced less scattering at longer wavelengths and yielded marginally larger residuals; however, the errors in the K-band are large and the constraints are weak given that very little of the reflectance in the H-band originates from the same aerosols probed in the K-band. By including shorter wavelength observations, Irwin et al (2015) had found the scattering cross-sections of aerosols at these heights diminished more strongly in wavelength, consistent with mean particle sizes of only 0.1µm. As Fig.16 shows, using a refractive index suggested by Irwin et al (2015) for the tropospheric cloud, the wavelength dependence across the K-band alone is similar in shape for both particle sizes, with a magnitude that may be shifted based on other factors folded into the effective scattering optical depth; with the H-band dominated by scattering from different depths, our solutions are degenerate among different combinations of scattering parameters. Therefore, based on expected trends across the K-band alone, our data can only be deemed consistent with particle radii less than a few microns. This conclusion holds if a refractive index of ~1.26 + i0.01 (Martonchik and Orton, 1994), appropriate for pure methane, is used instead.

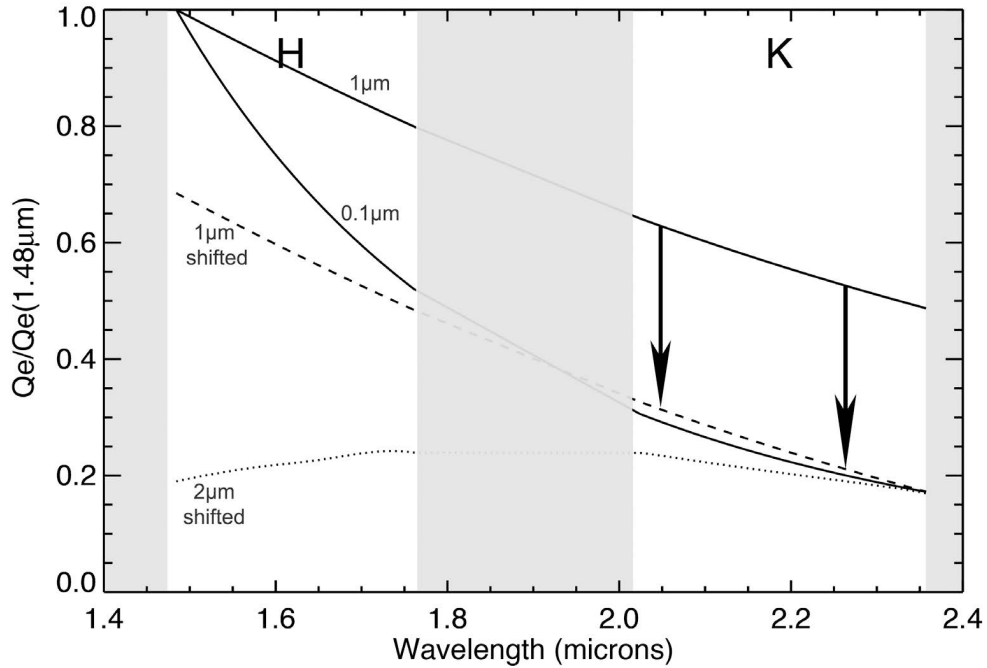

Figure 16. Aerosol extinction efficiency ratios versus wavelength for several different mean effective particle radii. The extinction efficiency at each wavelength is expressed as a ratio to the extinction at 1.48μm. The shape of the curves in the K-band are similar for different particle sizes and thus are only weakly constraining in the absence of reflection in the H-band, which is dominated by scattering from greater pressures. Calculations are done assuming Mie theory and a refractive index of 1.38 + i0.075, as found appropriate for the lower tropospheric by Irwin et al. (2015).

*4.3. Latitudinal and Time Variation*

Retrieved aerosol scattering profiles were plotted as a function of pressure and latitude for each observation. With the slit aligned roughly north-south at the central meridian, these data provide meridional cross sections for given times and longitudes. Figure 17 shows annual averages in scattering optical depths from 2001, 2002, 2004, 2005, and 2007 computed first assuming the Herschel/PACS methane profile and then the Spitzer methane profile; results retrieved using the free methane mole fraction are similar to those found using the Spitzer methane abundance, and hence are omitted for clarity. As shown for the mean 1-D profiles, the choice of lower stratospheric methane profile only alters the retrieved optical thickness of aerosols in the stratospheric haze layer. The choice of the deep methane profile, however, strongly affects the retrieved deep cloud optical thickness as discussed. We assume depletion towards the poles and hemispheric symmetry, but any unaccounted variation would lead to errors in the retrieved aerosol optical depths and inferred trends.

In general, these cross sections are similar in appearance to what Irwin et al. (2009) had produced using only H-band observations from 2006-2008; their results feature an aerosol layer extending over most latitudes at 2-4 bars with peak values of 0.3-0.4 optical depth per bar in the southern hemisphere–similar though somewhat deeper than our retrieved values. In all cases, the cross sections display significant trends in the lower layer scattering optical depth over latitude and time. The 2001 data shows greater scattering optical depths towards the southern latitudes, peaking around 50° S latitude, and decreasing towards the pole and northern hemisphere. The peak in aerosols corresponds to the latitude of the bright south polar collar. South of 50° S, the height of the lower cloud layer gradually drops, with the peak concentration located several hundred millibars deeper relative at lower latitudes. This appears to be a direct consequence of the assumed latitudinal trend in the deep methane abundance, which gradually places the transition to higher mole fractions at higher pressures poleward of 45°S based on analysis over a greater spectral range (Karkoschka and Tomasko, 2011), as opposed to an edge-effect. If we use a methane model that is constant in latitude, this trend is no longer present and we instead see a trend of clouds becoming optically thinner and slightly higher towards the pole, similar to what Irwin et al (2010) had found in 2006-2008 observations. While this feature hence entirely depends on the assumed deep methane model, this combination of deeper cloud and diminished methane provides a significantly better fit to the data than constant methane and constant cloud height, as shown by the goodness of the fits (Fig 11), and we therefore interpret it as a real feature.

North of the equator, the $\tau_s$ bar$^{-1}$ drops by an order of magnitude. As the years progressed towards equinox, the aerosol scattering optical depths decreased over most of the disk, as shown in the mean profiles of Figure 15, but most dramatically towards southern high latitudes. As the south polar collar diminishes over the period, a somewhat more hemispherically symmetric cloud layer is seen by 2007, but with the greatest optical depths still found in southern hemisphere. This is generally consistent though perhaps less pronounced than the asymmetry reported by Irwin et al., (2009) from 2007 observations (their figure 9). Their cloud has a similar optical depth per bar (<0.4 with a $\varpi_0$ =.75), but peaks and extends somewhat deeper between 2-8 bars with their assumed 4% deep-methane mole fraction. Sromovsky et al. (2007) similarly found the peak in cloud reflectivity around 45˚S in 2004, and the emergence of a bright region around 45˚ N.

The changes in time and latitude can be seen clearly in the total vertically integrated scattering optical depths shown in Fig. 18. Over the six-year period, the total integrated scattering dropped by more than a factor of three at high southern latitudes, but slightly increased at northern latitudes.

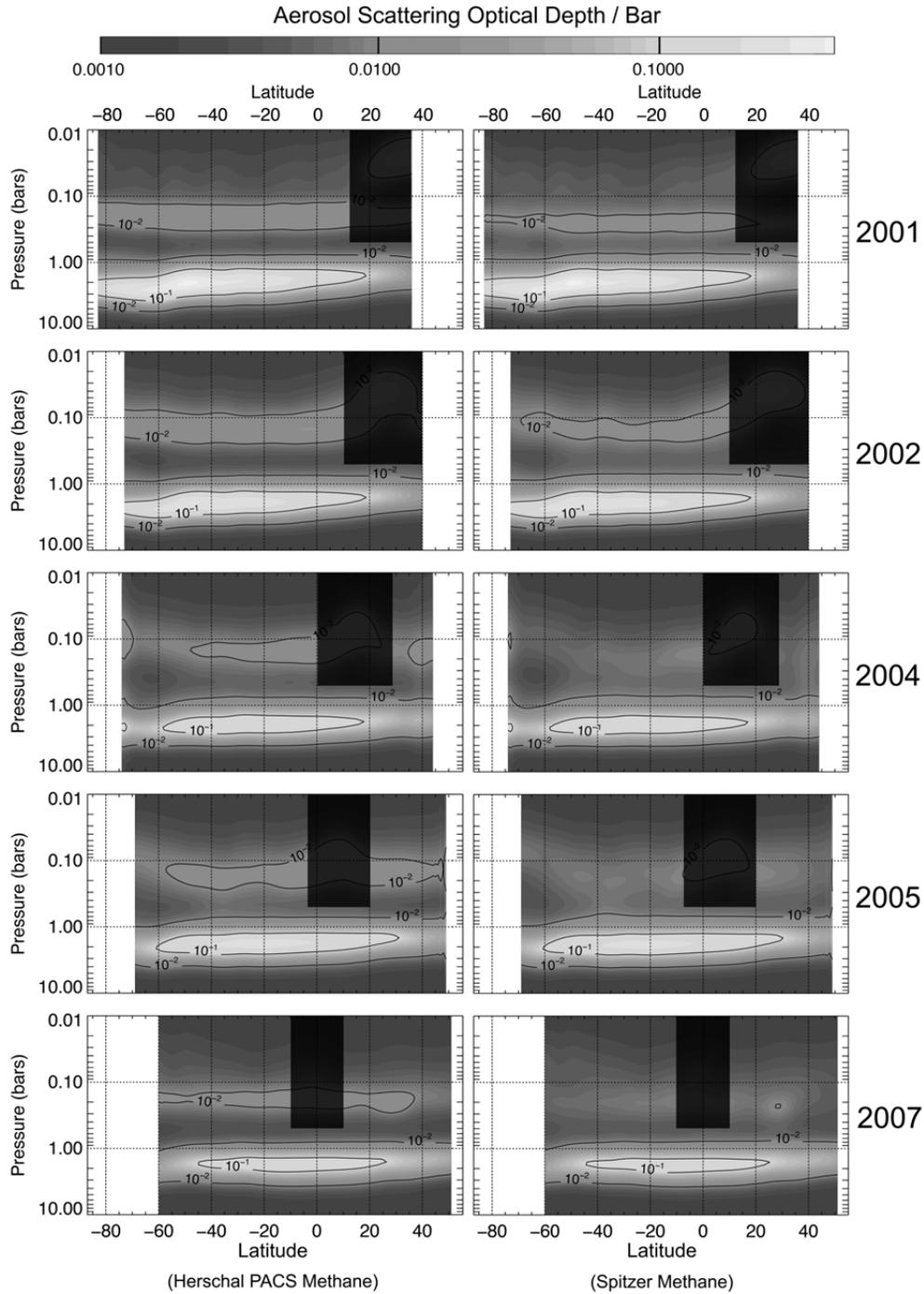

Figure 17. Aerosol scattering optical depth per bar are plotted as function of pressure and latitude averaged over each year for two assumed methane models: (left) Herschal/PACS and (right) Spitzer. Regions dominated by the ring reflectance in the K-band are blackened. There is a significant reduction in the $\tau_s$ bar$^{-1}$ with time at all latitudes but particularly southern latitudes, resulting in a more symmetric pattern by 2007.

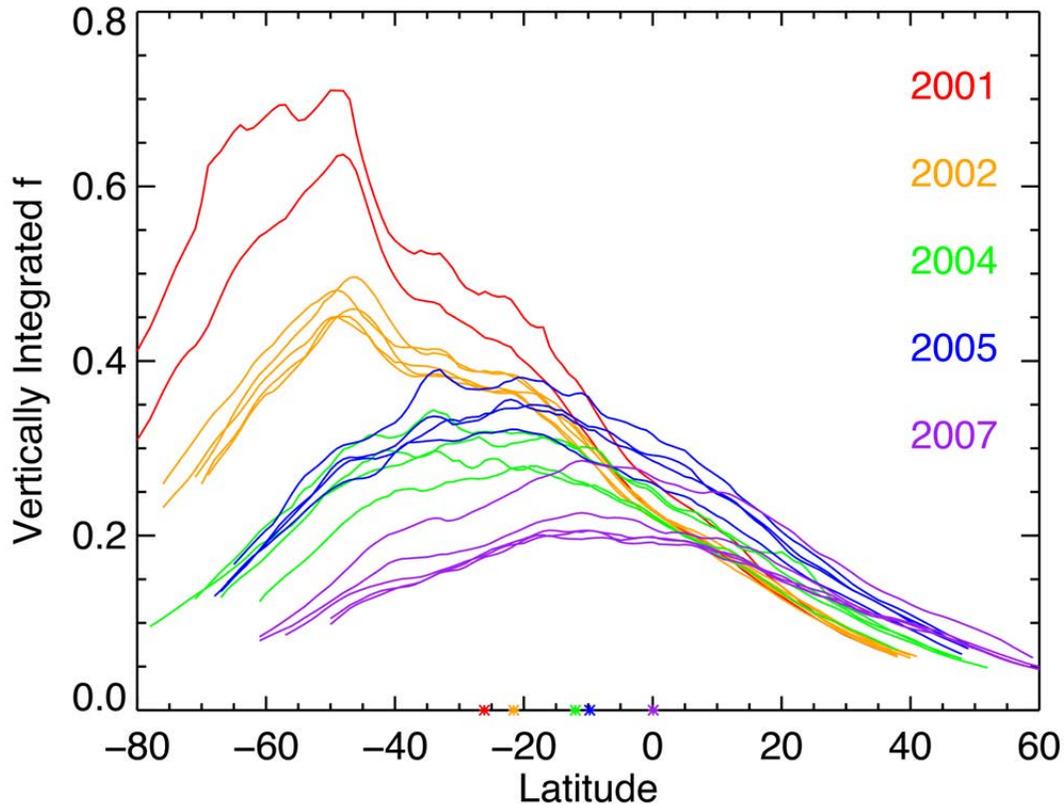

Figure 18. Vertically integrated scattering optical depths versus latitude for each of the observations, color-coded to indicate the year. Colored asterisks on the abscissa indicate the sub-solar latitude for each year. The data clearly shows a trend of diminishing scattering optical depth at southern latitudes progressing from 2001 to 2007.

When treated as a free parameter, the tropopause methane mole fraction showed no obvious trend or significant variation with latitude or time over the same period given an uncertainty of a factor of two in the retrieved values. The deep methane was assumed to have hemispheric symmetry and follow the trends discussed.

Evaluated Chi-squares were significantly less in the northern hemisphere than in the southern hemisphere (by as much as a factor of two). Since the northern hemisphere had a greater reflectance and aerosol opacity, the trend in Chi-squares likely indicated a limitation in our single-scatter modeling of the aerosol scattering properties or gaseous opacities in the brighter regions of the H-band as evident in the residuals of Fig. 14.

*4.4 Discrete Cloud Feature*

Occasional discrete cloud features can be seen throughout the data. A lone discrete feature appears to accompany the south polar zone in the H-band images from 2004 and 2005, and a number of discrete clouds can be seen at mid-northern latitudes (Figure 2). The most prominent discrete cloud feature recorded in our data was observed on August 17, 2007 at 30° N latitude and 225° E longitude. The feature was roughly 10 pixels or ~6,000 km across (FWHM ~4,000 km) and is seen as increased reflectance in both H and K bands (see Figure 18). The feature is likely the same as that captured in observations by Sromovsky et al., (2009) (including one observation only 49 hours after our observation) and Irwin et al., (2009). The extent of the feature is somewhat artificially exaggerated due to limitations in atmospheric seeing and rotation of the disk, while the intensity of the light is potentially diminished by 5-10% as suggested by deconvolutions of the point-spread function. Nonetheless, the feature reflectivity in the H-band was measured with an I/F of 0.0101—roughly 1.8 times brighter than nearby longitude. In the K-band, the cloud feature had an I/F of about $1 \times 10^{-4}$, also about 1.8 times brighter than the nearby longitudes. The K-band increase is unique within the data and required the feature to have existed higher in the atmosphere than other clouds. Spectra were fortunately obtained for the feature and adjacent regions of the atmosphere. As Figure 19 shows, enhanced reflectance was observed at wavelengths corresponding to moderate to strong methane absorption features in the H-band as well as the near side of the K-band.

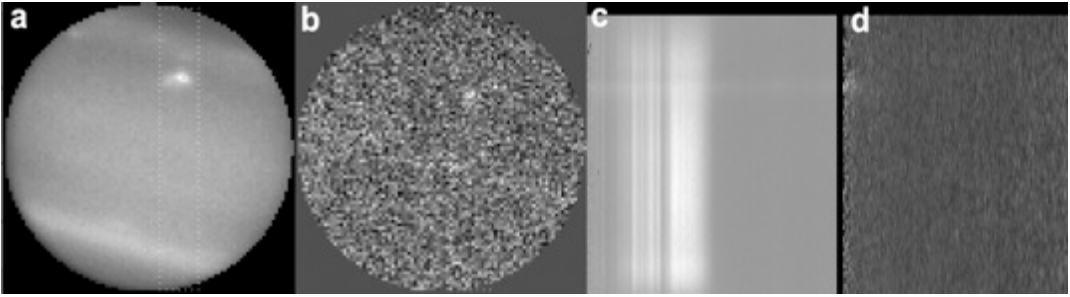

Figure 18. Images and compressed spectral images for the discrete cloud feature, as observed on August 17, 2007. The feature is seen as a bright spot just right of center in the upper half of the images. a) The H-band image. b) The K-band image. c) The H-band spectral image, with the vertical dimension corresponding to the range of latitudes covered by the slit, and horizontal corresponding to the range of wavelengths dispersed by the H-grism. The cloud feature increased reflectance across the entire H-band (1.49-1.78 μm). The image was compressed in the horizontal dimension for display. d) The equivalent K-band spectral image, showing just a hint of brightness on the left, corresponding to wavelengths ranging from 2.03-2.08 μm.

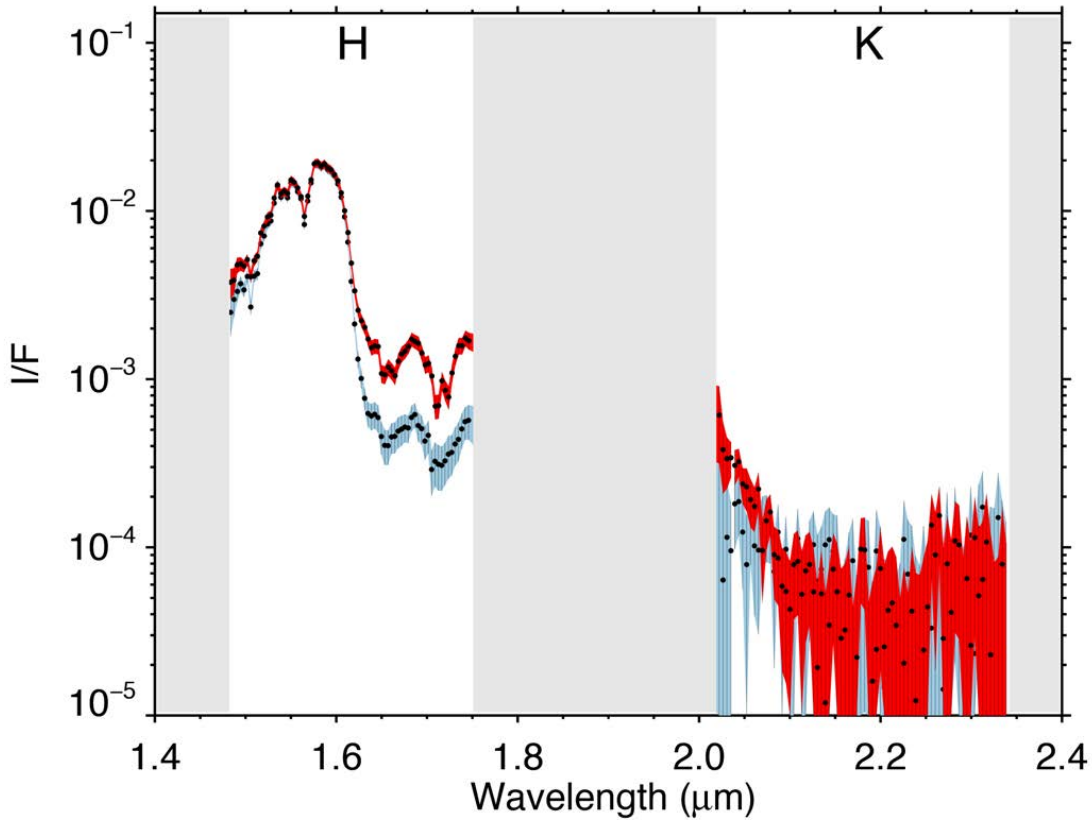

Figure 19. Spectra obtained for the discrete cloud feature seen in 2007 and shown in Fig. 18 (dark/red) and adjacent longitude (light/blue), along with error bars. Enhanced reflectance was observed at wavelengths corresponding to moderate to strong methane

absorption features in the H-band as well as the hydrogen absorption in the near side of the K-band.

The full latitudinal cross section retrievals suggests the cloud may be seen as an enhancement in a more tenuous and uniform layer between 100 and 300 mb with no significant change in the 100-mbar methane mole fraction (Figure 20).

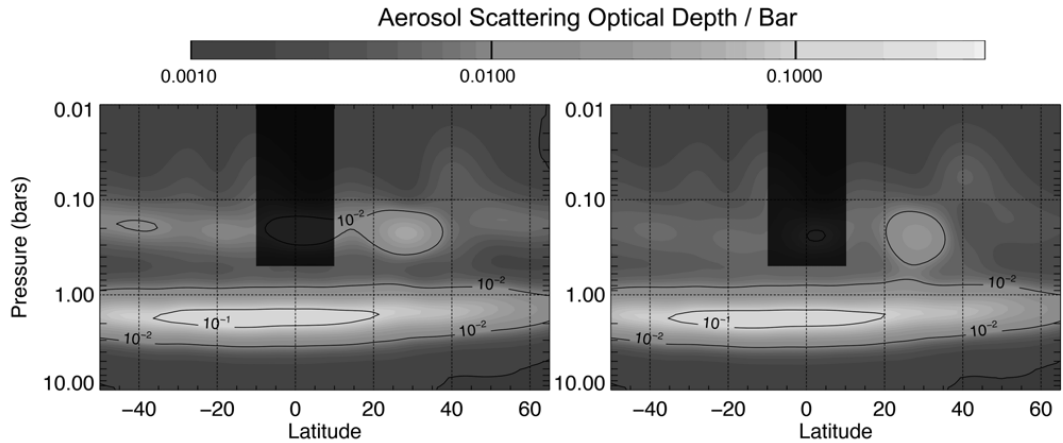

Figure 20. Latitudinal cross section retrieved from August 17, 2007 observations for the two preferred methane profiles: (left) Herschel/PACS and (right) Spitzer. The bright cloud feature (see Fig 18; Fig 19) appears as an enhancement of aerosol scattering optical depth at roughly 250 mbar and 30°N latitude.

Retrievals reveal that the observed feature corresponds to an increased aerosol scattering optical depth centered at 240 ± 60 mbar assuming the Hershcal/PACS methane value, or 262 ± 65 mbar assuming the Spitzer value. Vertical profiles show an additive enhancement of 0.027 or 0.054 $\tau_s$/bar, respectively, depending on the assumed methane profile, relative to the surrounding atmosphere (Figure 21). Vertically integrated between 100 mb and 900 mb, both these values correspond to a 0.007 ± 0.002 additive increase in scattering optical depth relative to adjacent locations.

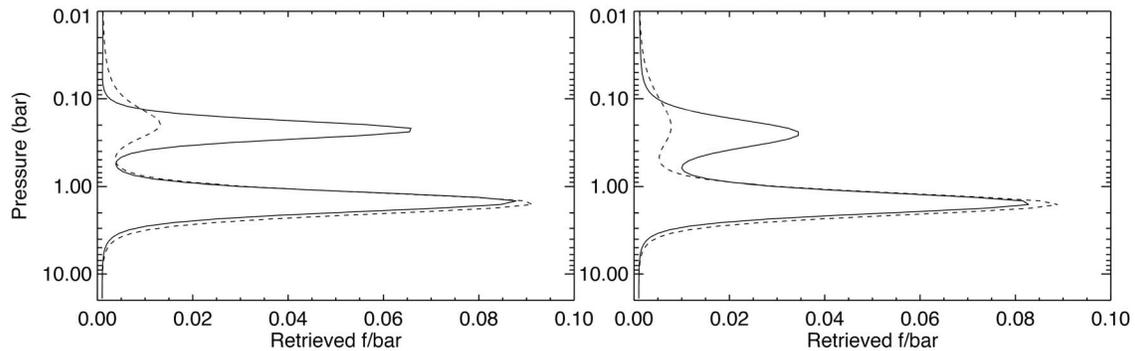

Figure 21. Plots show the retrieved vertical profiles of the scattering optical depth per bar for the 30° N discrete cloud feature observed in 2007 (Fig 18) and adjacent locations for each methane model: (left) Herschel/PACS and (right) Spitzer methane profile. The solid line depicts the retrievals centered at the latitude and longitude of the discrete cloud; the dashed lines show the average of the retrieved values for immediately adjacent latitudes and longitudes.

## *5. Discussion*

### *5.1 Cloud Structure and Seasonal Changes*

The present study adds to a large number of studies that have examined cloud structure and seasonal changes on Uranus using near-IR spectra (e.g. Baines and Bergstralh, 1986; Sromovsky and Fry, 2007; Sromovsky et al., 2011; Irwin et al., 2009; Irwin et al 2010; Irwin et al., 2012a; Tice et al., 2013; Irwin et al., 2015 to name a few). While the need for at least two vertical aerosol layers is consistent with an overwhelming majority of these studies, differences in the precise number and nature of parameterized cloud layers certainly exist. Models have included the use of discrete, compact layers (Fink and Larson, 1979; Sromovsky et al., 2006, Irwin et al, 2012; Tice et al., 2013), extended diffuse layers (Karkoshcka and Tomasko, 2009), and combinations of the two (Sromovsky et al., 2012; Sromovsky et al., 2014; Irwin et al., 2015; de Kleer et al., 2015). For example, Sromovsky et al. (2014) used a five-layer model, with three compact aerosol layers, representative of condensate clouds, beneath two more vertically extended but optically thin layers, more representative of hazes. In our approach, the nature of our retrieval method and the intrinsic information content of our data yield solutions with smooth, broader peaks in place of sharp, compact layers, as discussed in Sec 3.3.1. And though restricted in form by the imposed smoothness constraints, our solutions were not forced to include any particular number of vertical layers. As test cases showed, three broadly separate aerosol layers between roughly 100 mbar and 3 bars should be successfully retrieved using our approach (see Fig. 9), but we note that the vertical spacing of the layers is important. Multiple, closely spaced, compact aerosol layers, as in the Sromovsky et al. (2014) five-layer model, would simply be unresolvable in our retrieved profiles. Likewise, layers deeper than a few bars (like the bottom cloud in the same five-layer model) would fall beneath the range of our observations and hence be missed. Therefore, our results do not rule out the possibility for more than two aerosol layers in the upper troposphere–our data simply do not demand more. Despite differences in applied methods, this conclusion is nonetheless reassuringly consistent with a majority of the previous studies probing similar heights.

The apparent relative minimum in aerosol scattering between the two layers found in the present retrievals is qualitatively similar to results found for Jupiter and Saturn (Banfield et al., 1998a, Stam et al., 2001). This may indicate a process of photochemical aerosol production in a stable layer below which aerosols are cleared by gravitational fallout, as explored by Banfield et al., (1998a), or simply evaporated and diluted below the source.

Though inferred in a majority of near IR studies, the roughly 2-3 bar pressure of the deeper cloud layer is at odds with the 1.2 bar height inferred from Voyager Radio occultation analysis (Lindal et al. 1987). This discrepancy was thoroughly investigated by Sromovsky et al., (2011), who concluded that most modelers were assuming deficient methane mixing ratios—which moved the clouds deeper—and that the lower layer may be composed of multiple layers—namely a very thin 1.2 bar methane cloud and a deeper hydrogen sulfide cloud—lumped together in retrievals. Determining whether this layer is composed of a single convective condensate cloud layer, multiple condensate cloud layers, or a semi-infinite haze is beyond the constraints of this present data and analysis; however, the changes in its properties over latitude and time may provide insight into its nature.

While we are still far from understanding Uranus' global circulation patterns, Sromovsky et al. (2014) suggested a scenario in which three layers of circulation cells lead to different cloud layers at different heights and latitudes. In such a scenario, a roughly 1-bar methane cloud would form at the equator and extend 45° north and south; below this layer, a 1.5-bar hydrogen sulfide cloud would exist at mid-latitudes in an upwelling branch of a deeper circulation cell. Though our retrieved cloud heights are somewhat deeper, the qualitative picture may apply. In this context, perhaps the retrieved change in cloud height at 50° marks where the methane cloud ceases, leaving a slightly deeper $H_2S$ cloud exposed. If a transition is present, it may be related to the transition in opacity at 45° at much greater depths suggested by observed microwave brightness (Hofstadter and Butler, 2003), which increased in intensity following the previous solstice. However, the retrieved change in cloud height is small and there is no other direct evidence to suggest a change in cloud composition; as such, relations between the cloud distributions and any hypothetical circulation patterns clearly remain speculative and requires further research.

The dramatic change in the reflectance and total scattering optical depth from 2001 to 2007 (see Fig. 18) is consistent with other studies using observations covering different periods between 2004 and 2011 (Irwin et al., 2009, 2010, 2012; Sromovsky et al., 2009). Analysis of observations from 2010 and 2011 by Irwin et al. (2012b) and de Kleer et al. (2015) show that the northern hemisphere continued to become more reflective as compared to equivalent southern latitudes. The retrievals show that the observed trend is due to changes at the ~2 bar layer.

The scattering phase angle changed by less than a degree between 2001 and 2007, so the observed changes are truly intrinsic changes in the particle scattering properties, the assumed methane abundance, or the aerosol abundance. Since $\tau_s$ is a product of the scattering parameters, the observed change in reflectance may potentially be due to changes in any of its component factors. The scattering phase angle or single scattering albedos may potentially change with changing particle sizes or aerosol chemistry; while the data showed no significant change in the extinction efficiency behavior and related particle size, only a small range in wavelengths were sensitive to the 2 bar cloud layer associated with the hydrogen sulfide cloud. Temporal changes in the single scattering albedo are also possible.

Such rapid response in the aerosol abundance due to seasonal changes is surprising given the potentially long radiative and dynamical timescales. Conrath et al. (1998) suggested radiative and dynamical time constants of 130 and 700 years at 300 mbar based on temperature and para-hydrogen fields inferred from Voyager IRIS observations. As Irwin et al. (2009) notes, if radiative timescales are longer than a year, the observed uniform thermal emission implies that Uranus' atmospheric circulation is capable of fully redistributing the solar heating away from the substellar point. Dynamical time scales are presumably shorter in deeper, convective regions of the atmosphere, and changes in the aerosol optical thickness may be a result of weakening vertical mixing or upwelling velocities in the southern hemisphere. In context of the hypothesized circulation model discussed above, this would be most directly attributed to weakening of the mid-level circulation cell and the associated hydrogen sulfide cloud.

Alternatively, the rapid response may suggest a direct insolation or photochemical component. Studies of Saturn's atmosphere have shown that the hazes have clear seasonal trends. Pérez-Hoyos et al. (2005) noted that optical thicknesses of Saturn's stratospheric hazes at visible wavelengths increased with increasing insolation, which makes sense if photochemical production was the source; however, the deep tropospheric hazes *decreased* with increasing insolation (Pérez-Hoyos et al., 2005). Conversely at 5 μm, the tropospheric haze and clouds appear optically thicker in the summer hemisphere (Fletcher et al., 2011). In Saturn's case, the explanation has been attributed to increases in the particle size and the relative extinction efficiencies. Particle sizes were found to be larger in the summer than the winter hemisphere by a factor of 2 to 4 (Karkoschka and Tomasko, 2005). In the case of Uranus, a change in effective mean particle sizes may largely explain the significant change in aerosol scattering optical depth at 1.48 microns with little to no seasonal lag. For example, a factor of five reduction in the scattering extinction efficiency could be accomplished by reducing the scattering size parameter, $2\pi a/\lambda$ (where $a$ is the effective mean particle radius), from 6 to 2 in a cloud with an effective size variance of 0.1 or less (Hansen and Travis, 1974). At 1.48 μm, this translates to a change in particle radius from about 1.4 μm to 0.5 μm. Irwin et al (2015) reported a particle size of 0.86 ± 0.04 μm at the 2 bar level in analysis of 2009 observations, but most modelers have assumed particles were roughly 1 μm in radius in the past due to the poor leverage afforded in the narrow deeply penetrating passbands. Note, however, that changes in the particle size alone would be expected to change the extinction efficiency and reflectance at other wavelengths. All else being equal, the change proposed above would lead to a significant increase in scattering optical depth in visible wavelengths, which is inconsistent with the slight, gradual slight darkening observed in southern latitudes over the late 1990s (Rages, 2000; Karkoschka, 2001); however, comparisons between wavelengths may also be complicated by changes in normalized scattering phase function and single scattering albedo. It remains possible that a change in particle size and scattering properties could contribute to the observed seasonal changes, suggesting the need for a more thorough modeling of the scattering parameters (e.g. Irwin et al, 2015).

Another alternative explanation may instead look to changes in the methane abundance, on which the retrieved scattering optical depths are dependent. If our assumption of hemispheric symmetry and latitudinal dependence is incorrect, the retrieved optical depths may show false trends. It is plausible that a decrease in the southern hemisphere methane abundance would reduce the amount of absorption and cause a constant aerosol abundance to produce greater reflectance. Simultaneous constraints on the aerosols scattering properties and methane over time would be needed to attribute the source of the change with some certainty.  Progress in this regard could be achieved with a series of spectral observations covering wavelengths from the visible to the K-band, including the critical hydrogen dominated window at 825 nm (Karkoschka and Tomasko, 2011; Sromovsky et al., 2014).  These observations would likely need to be acquired annually for several years or more to show possible changes in methane and aerosols associated with seasonal forcing.  Concurrent characterization of the aerosol scattering properties will likely require high-resolution images in a number of visible and near-IR filters with accurate photometric measurements of the disk's center-to-limb brightness.

*5.2 Discrete Cloud Formation*

Several observers have noted discrete cloud features on Uranus over the past decades. Eight discrete clouds were seen in Voyager images (Smith et al. 1986), and Karkoschka (1998, 2001) summarized activity including several dozen spots between 1994 and 2000 seen in HST images.  A number of particularly bright clouds have been observed at mid-Northern latitudes in 1999, 2004, 2005, 2007, 2011, and 2014 (Sromovsky et al. 2000; Sromovsky and Fry, 2004; Sromovsky et al., 2007; de Pater et al., 2014).  A majority of the brightest clouds were observed around 30° N and appeared to persist on timescales of at least days and possibly years.  The August 2005 feature was the brightest observed up to that time, and was centered at 30.2°N (Sromovsky et al., 2007).  Observations a year later showed features at roughly the same latitude, but they also revealed an accompanying dark spot.   June 6, 2007 Keck II images (Sromovsky et al., 2009, 2012) once again showed a bright complex at roughly the same latitude apparently associated with a dark spot. Sromovsky et al., (2009) analyzed the motions of what they described as a bright complex at 30°N between June and August of 2007. They determined the feature exhibited oscillations in latitude and longitude while drifting westward; they also noted a possible accompanying dark spot. On August 19th they reported that the bright cloud complex was centered at 30.28°N latitude and 216° E longitude. Though they did not determine vertical height, the bright K-band observations suggested it was very high in the atmosphere. Irwin et al. (2009) also observed a cloud at 30°N in June and July of the same year, and reported its cloud top at the 100-200 mbar level. The feature we observed on August 17, 2007 at 30°N latitude, 225°E longitude, is almost certainly the same cloud complex observed by Sromovsky et al (2009) and likely the same reported by Irwin et al (2009).

For several similar features, radiative transfer analysis showed that such clouds reached heights of roughly 500 to 350 mbar (Sromovsky et al., 2007). At these heights, the atmosphere is very stable to convection. Aerosols could potentially be lofted to these heights through deep convection if updrafts are sufficiently energetic to overcome the stability (e.g. Lunine and Hunten, 1989; Stoker and Toon, 1989). Indeed, for the 2007 feature discussed above, Irwin et al (2009) suggested its height indicated a very vigorous storm penetrating to the stratosphere. Such convection would have to originate at great depths where condensation of water and other deep constituents can contribute to latent heating and drying of the parcel. With positive buoyancy, the parcel would then need to accumulate enough kinetic energy over its ascent to overcome the large stability at above the ~1 bar level. Such convection evidently occurs, as shown by the vertical structure inferred by Irwin et al. (2016) for clouds in 2014, with the cloud extending vertically over several bars. In our case, the retrievals do not necessarily require a vertically extended cloud to match the observed spectral reflectance. Indeed, given the inferred aerosol enhancement, the parcel would also need to be very dry by the time it reaches the ~250 mbar height in order to produce a number density of aerosols consistent with the observed scattering optical depths. The thin vertical extent and optical thinness may also suggest the cloud complex itself is more akin to a cap or pileus cloud, forming from ambient gas above– and detached from –an energetic convective plume beneath. The gas within plume would apparently need to be very dry or produce a very absorbing condensate in order to contribute so little aerosol reflectance beneath the observed cap.

Alternatively, the cloud features could be formed by other means than convection. The apparent longevity, oscillatory motion, and possibly accompanying dark spot all suggest these cloud complexes may be associated with vortex circulations. This would be akin to the bright companion clouds seen over the dark spot on Neptune in Voyager 2 images (Smith et al., 1989) and have been thought of as somewhat like terrestrial orographic clouds. Neptune's companion clouds have been modeled as flows along the surfaces of constant density (isentropic surfaces) rising over perturbations due to local temperature anomalies (Stratman et al., 2001). Clouds formed by this process should be very thin and composed of small, relatively uniform particles sizes given the modest, laminar lift and the very limited condensable methane content available at their inferred altitudes. The retrieved optical thickness of this cloud can be used to help evaluate the consistency of the observations with formation models.

From the observed spectra, the retrieved optical thickness and vertical extent of the cloud was inferred. By approximating aerosols as spherical particles with scattering cross sections of $\sigma = \pi a^2 Q_s$, then the optical thickness per unit area can be roughly converted to a mass per unit area:

$$M = 4 a \rho \tau_s / 3 \, PF \, Q_s$$

Where $M$ is the total aerosol mass per unit area, $a$ is the effective particle radius, $\rho$ is the particle density, $PF$ is the normalized scattering phase function, and $Q_s$ is the scattering extinction efficiency. The scattering extinction efficiency is the product of the total extinction efficiency, $Q_{ext}$, and the single scattering albedo, $\varpi_0$. To evaluate this expression, a particle radius of ~1μm can be taken from our best fitting retrievals for this location. Note that this size is an order of magnitude larger than the typical ~0.1μm aerosols sizes reported in the hazes at similar heights, though similar in size to aerosols found in deeper tropospheric clouds (Tice et al., 2013; Irwin et al, 2015); this size discrepancy may be reconciled if this discrete cloud feature has a composition and genesis that is more akin to the deeper condensate clouds than the stable hazes with potential photochemical origins.

If we assume the cloud aerosols have scattering properties similar to the deeper condensate clouds, then we may estimate a single scattering albedo of ~0.7 based on results of Tice et al. (2013) and Irwin et al. (2015). If a refractive index of 1.38 + i0.075 is chosen, again following Irwin et al. (2015) for a lower tropospheric cloud, then the above single scattering albedo would yield a Mie scattering extinction efficiency, $Q_s$, with an average value of ~2.5 over the range of wavelengths significantly brightened by the cloud. Tice et al. (2013) found a normalized Henyey-Greenstein (HG) scattering phase function with a forward asymmetry parameter of 0.7 to provide a good fit, but given the low optical depths and viewing geometry, the backscattering parameter of a two-term HG phase function would be more relevant to our calculation. A two-term HG phase function with a backscattering parameter of g~0.3 has been found to provide good fits for aerosols of roughly similar size, indices of refraction, and forward scattering asymmetry in Saturn's atmosphere (Tomasko and Doose, 1984; Pérez-Hoyos et al., 2005; Roman et al., 2013; Pérez-Hoyos et al., 2016); assuming a roughly similar phase function (*PF*) would apply, evaluated at a scattering angle of 179°, we let $PF(179°) \approx 0.4$. Finally, solid methane at a 60 K has a density of 0.50 g cm$^3$, while liquid droplets have a density of 0.42 g/cm$^3$ (Bol'shutkin et al.,1971; Roe and Grundy, 2012); the density may be lower if the aerosols are icy with significant porosity. Assuming a density of 0.5 g/cm$^3$, the retrieved total scattering optical for the layer, $\tau_s = (7.2 \pm 1.5) \times 10^{-3}$, yields a mass of $(8.6 \pm 1.7) \times 10^{-3}$ g/m$^2$. Alternatively, if we assume the aerosols are more akin to typically haze particles, then according to Tice et al. (2013) and Irwin et al. (2015) a radius of 0.1μm and $\varpi_0 \approx 1$ would be more appropriate. In this case, the size parameter and resulting scattering extinction efficiency become small (i.e. $Q_s \approx 0.2$), and the scattering phase function becomes more isotropic with $PF(179°) \approx 0.7$, yielding a mass per unit area of $(3.0 \pm 0.6) \times 10^{-3}$ g/m$^2$.

The total mass of condensate of the observed cloud can then be related to the amount of condensate expected by condensation of a rising parcel of atmosphere. Using a classical parcel theory description, the parcel begins with a given methane mixing ratio and temperature, typically taken to be equal to the surrounding ambient atmosphere; as it rises, the parcel does work on the surroundings and consequently cools as it loses energy. If the equilibrium vapor pressure of methane in the parcel exceeds the temperature dependent saturation vapor pressure, the excess vapor condenses out forming aerosol.

The aerosol may then be left at the height it condensed (a pseudoadiabatic process) or taken along with the parcel (a reversible adiabatic process), or a combination of the two as is typical in terrestrial cases. Here it is assumed that the aerosol is pseudoadiabatically left behind for simplicity, but given the meager vertical distances and mass condensate, the choice of process is inconsequential to the result. The buoyancy of the parcel can be evaluated at each step of its rise, and the energy required or released by the ascent can be tallied. Additionally, complications can be added by including entrainment and supersaturation; for these calculations, entrainment is parameterized following Stoker and Toon (1989) and condensation is delayed to allow a supersaturation of 5% based on terrestrial values.

This process can be plotted using a classic thermodynamic diagram as shown in Figures 22 and 23. The retrievals place constraints on the cloud height, total aerosol mass per unit area of condensate, and the methane relative humidity of the environment. Through the simplified thermodynamics of parcel theory, these retrieved parameters translate to constraints upon the initial conditions of the parcel. If the parcel is assumed to have an initial methane abundance and temperature set by the retrieved methane profile and assumed temperature profile, then the condensation height is unique to a certain starting pressure, depending on the methane profile, previously stated scattering assumptions, and assumed thermodynamic properties. For the following calculations, we use an Antione equation for methane vapor pressure (with values from Prydz, Rolf, and Goodwin, 1972) and a latent heat of sublimation of 605 J/g (Stephenson and Malanowski, 1987) for methane, though we find our conclusions are unchanged for the range of published values for the latter.

If we assume the methane profile inferred from Spitzer, a lifted condensation level (LCL) at the discrete cloud's inferred pressure of 260 mbar (Figs 18-21) would require a parcel to begin its ascent at 340 mbar, where the ambient relative humidity of methane is ~19%. (Figs 18-21). The amount of condensate produced depends on how far the parcel is lifted beyond the condensation level. To be self-consistent with the values inferred from the retrievals, if we assume 0.1μm radii particles, the parcel would have to rise to 256 mbar, just 4 mbar beyond the LCL, in order to condense the $3 \times 10^{-3}$ g/m$^2$ and necessary to yield an optical thickness of ~0.007. Lifting an additional 4 mbar would be necessary to produce the additional mass needed for 1μm particles to yield the same optical thickness. An 84 mbar change equates to roughly 1/3 of a pressure scale height (H~21.77 km) and a temperature drop of nearly 6 K. The atmosphere is stable at this height, so even with the relatively small contribution from the latent heat of condensation, the lift across a vertical density gradient would require ~50 J of energy per kilogram of atmosphere (i.e. the convective inhibition (CIN)). If this 84 mbar ascent was accomplished by initial momentum alone, as from the impulse of an underlying convective plume, then equating this energy to the amount of kinetic energy, necessary to overcome this potential energy yields a vertical speed of ~8 m sec$^{-1}$. Alternatively, if the parcel is thought to move along a sloped isentropic surface, this suggests that the anticyclone would have a horizontal temperature gradient comparable to the temperature drop (~6.0 K).

Smaller temperature drops and energetic barrieres can be accommodated if the parcel begins with a slightly greater methane mole fraction and at a slightly lower pressure. If we repeat this exercise for the more humid Herschel/PACS profile, we find that the parcel could achieve condensation at 240 mbar (the central pressure retrieved for using that methane model) by starting the parcel at 266 mbar, where the relative humidity would be ~58% (Fig 23). The parcel would only need to ascend less than 2 mbar beyond the condensation level to produce the required condensate mass when composed of 0.1μm particles (3 mbar if 1μm particles are assumed). The modest ~28 mbar ascent would only result in a 2 K temperature drop and require only 1.7 J/kg of energy to overcome the convective inhibition, corresponding to an initial vertical velocity of less than 2 m/s.

For comparison, the EPIC modeling of Neptune's companion clouds by Stratman et al. (2001) found that those clouds were best reproduced with a typical pressure drop of 3 mbar from the 76 mbar height, corresponding to half a kilometer, 4% of a scale height, and 1 K temperature drop. The difference could largely be explained by a relatively lower methane mixing ratio in the Uranus atmosphere, which requires a greater ascent prior to reaching condensation assuming the parcel begins at the ambient humidity. If the 200-400 mbar methane mole fraction were closer to saturation, the required ascent would be less. If the environment were assumed to be saturated in the above calculations, with no supersaturation necessary, the parcel would condense immediately upon rising, requiring just the 2-3 mbar ascent from 240 mb which corresponds to roughly 0.3 km in height, ~1% of a pressure scale height, and a 0.2 K temperature drop—somewhat more consistent with the Neptune modeling results.

Similar analysis may be applied to penetrative deep convection, but solutions require extreme initial conditions. The deep atmosphere is much richer in methane than the stratosphere, and so the mass of condensates quickly becomes very large. In order to produce the aerosol mass and height inferred from the retrieved scattering optical depths while still achieving enough buoyant energy to penetrate the stable upper troposphere, the rising parcel would have to begin extremely dry and warm relative to its surroundings. For example, a parcel beginning its ascent at 1 bar would need to be roughly 3 K warmer and a factor of $3 \times 10^{-3}$ dryer than the surrounding atmosphere in order to condense the same amount of aerosols at the observed heights. Conceivably, initial conditions could be more moderate if entrainment and mixing were much more vigorous than modeled here, but nevertheless, deep convection without significantly higher optical thickness appears unlikely. The retrieved scattering optical depths are thus more consistent with lift initiated near the observed cloud heights, as expected with companion clouds flowing over vortex circulations. With the retrieved methane abundances and assumed scattering properties, the vortex would be forcing lift from roughly 340 mbar up to 256 mbar in the dryer Spitzer case, or from 266 mbar to 238 mbar in the more humid Herschel/PACS case. While uncertainties in the parameters can produce a range of different values, the qualitative picture remains unchanged.

Interestingly, a different discrete cloud feature at mid-northern latitudes in 2011 data was analyzed in a separate study and inferred to have a greater vertical extent—extending from 1.3 bars up to the 0.5 bar level (de Kleer et al., 2015). The authors concluded that clouds were consistent with convective upwelling with particles being lofted high above the deep condensation level. Likewise, Irwin et al. (2016) found discrete cloud in 2014 extending vertically over several bars, tilted by sheering winds and more consistent with a convective storm. The existence of both convective and vortex companion clouds at similar latitudes is consistent with arguments that anticyclonic vortices form in regions of strong upwelling and divergence aloft (e.g. as discussed by de Pater et al., 2014). As shown here, a radiative transfer analysis can help to distinguish between potentially different cloud formation processes and reduce the ambiguity of observed features.

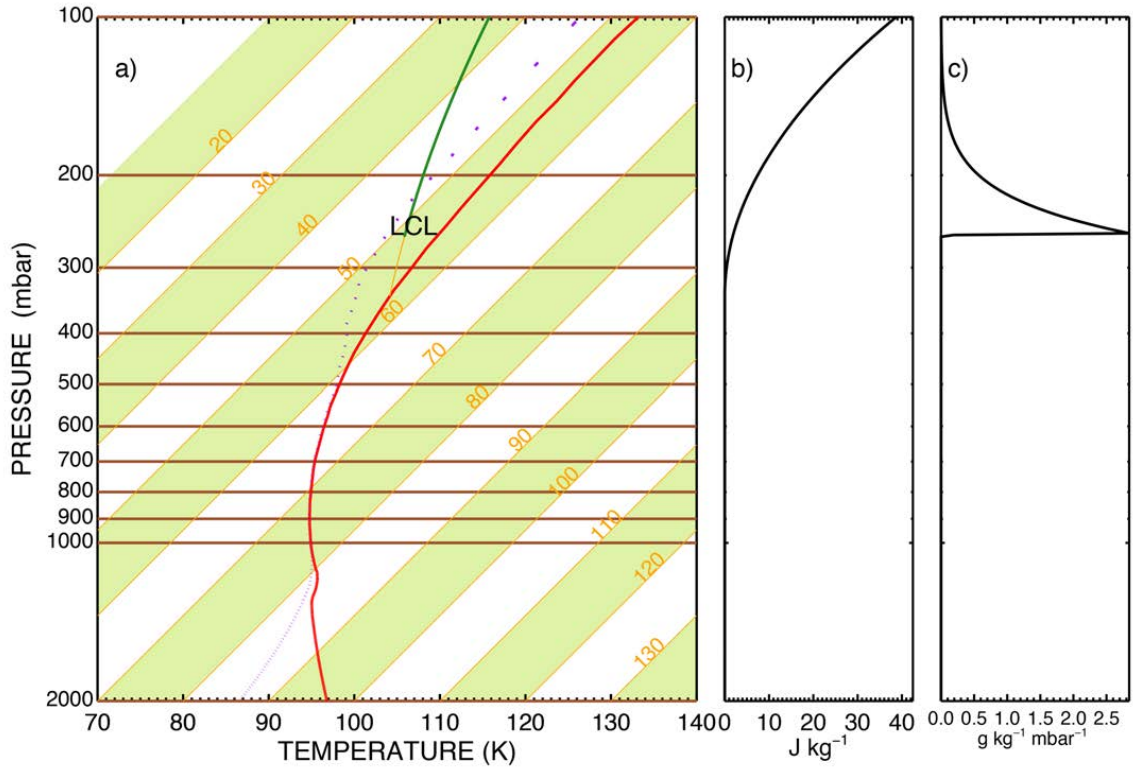

Figure 22. a) Thermodynamic diagram showing a hypothetical sounding for the Uranus atmosphere assuming the methane profile inferred from Spitzer. Temperatures are skewed at a 45° angle, increasing down and to the right. Pressure is on the y-axis, increasing down. The red line represents the assumed temperature profile ('model F' of Lindal et al. 1987). A purple dotted line represents the methane saturation temperature for the ambient atmosphere, not the parcel. The temperature of a parcel of air forced to ascend dry adiabatically from 340 mb is plotted in orange as it deviates from the ambient atmospheric profile; the line color becomes green when saturation in the parcel occurs (marked LCL for Lifting Condensation Level) and the parcel rises wet pseudo-adiabatically. b) The amount of energy (Joules) necessary to raise the negatively buoyant parcel (convective inhibition) per kilogram as a function of vertical pressure. c) The amount of condensate (grams per kilogram of atmosphere per mbar of ascent) produced as a function of pressure. The amount of ascent beyond the LCL determines how much condensate is produced.

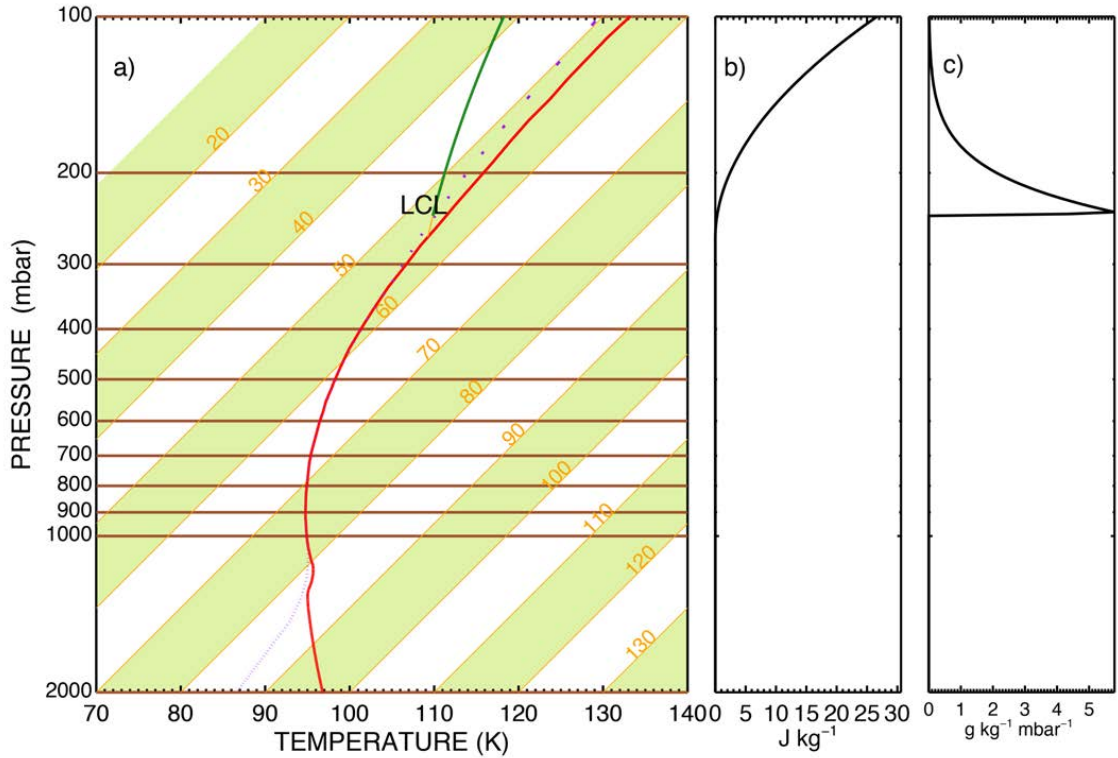

Figure 23) Same as Fig. 22, but now assuming the more humid Herschel/PACS methane profile and corresponding cloud level at 240 mbar. Less ascent is need to reach condensation and produce sufficient condensate relative to the dryer Spitzer methane profile.

## 6. Concluding Summary

Spatially resolved, Near IR spectra of Uranus were analyzed to infer the atmospheric distribution of aerosols and methane.

Aerosol scattering optical depths showed two distinct layers—an optically very thin haze near the tropopause and a deeper, thicker haze or cloud layer. The lower layer peaked at a pressure of ~1.94 bars (with a 10% uncertainty), and exhibited changes in reflectance with latitude and time. Scattering optical depths were found to be greatest in the southern hemisphere, centered near ~50° S in 2001. From 2001 to 2007, the total scattering optical depth of the deeper layer diminished by a factor of three or more, particularly at southern summer latitudes, resulting in a somewhat more symmetric but still skewed distribution. This analysis adds to a limited but growing body of observations documenting temporal changes on Uranus.

The changes in reflectance from 2001 to 2007 appeared to be linked to seasonal change as Uranus arrived at equinox, suggesting insolation strongly affects the aerosol distribution. Given potentially long radiative and dynamical time scales, the relatively rapid changes may possibly be attributed to changes in aerosol scattering properties or methane abundances, which were here assumed constant in time; however, the present data cannot distinguish between several potentially changing scattering properties and potential changes in aerosol or methane abundances. Recurring spectral observations that include the hydrogen-dominated window at 825 nm (Karkoschka and Tomasko, 2011; Sromovsky et al., 2014) appear necessary to separate the factors, along with detailed analysis of the disk center-to-limb brightness over a range of wavelengths and time.

The overall distributions of aerosols were consistent with—but not strongly discriminating among—speculative circulation models that include at least two layers of circulations cells at different latitudes and pressures. Such models may explain possible changes in cloud heights and reflectance seen at ~50˚S if the upper tropospheric methane abundance is assumed to diminish towards the poles.

Best fits to the data were indeed achieved with a methane distribution that decreased towards the poles at pressures >1 bar, consistent with the findings of recent studies (Karkoschka and Tomasko, 2011) and circulation models with downwelling at the poles. Given weak constraints on the methane mole fraction, we performed retrievals of aerosols under several different assumptions of the lower stratospheric/upper tropospheric methane abundance. When treated as a free parameter, the methane mole fraction at the tropopause was retrieved with an average value of $1.2 \times 10^{-5}$, significantly below saturation and consistent with Spitzer results (($1.6 +.2 /-.1) \times 10^{-5}$) and previous findings (Orton et al., 2014), but with an uncertainty as large as a factor of two given the possible degeneracies and lack of sensitivity. No significant variations in latitude were detected beyond our estimated uncertainties. Values based on the Spitzer and Herschel/PACS ($4.7 \times 10^{-5}$ at the 89 mbar) were also separately used for retrievals. Using the Spitzer value, we retrieved results similar to those found using the free-parameter for methane. The greater methane abundance in the Herschel/PACS case requires higher aerosol optical depths to match the observations, but the results were qualitatively similar. Potential temporal changes in the 100-mbar methane mole fractions have not been thoroughly investigated and thus require additional study.

Finally, a discrete cloud feature was analyzed. Despite the relative height and contrast, the cloud's observed reflectance, retrieved scattering optical depth, and inferred aerosol mass is very low, implying only modest lift. All together, the results favor the interpretation of this particular cloud as a bright companion cloud or complex of clouds associated with dark or unseen vortices.

Using Earth-based observations aided by adaptive optics, the approach taken above attempts to discriminate between potential cloud and haze formation dynamics using observations of reflectance, previous findings of the temperature and methane profiles, and a series of self-consistent assumptions made necessary by the lack of data. Stronger constraints on the methane distribution at the highest cloud levels are difficult given the very low aerosol optical depths and signal to noise ratio in the strongest methane bands. Likewise, greatly reduced uncertainties in the K-band reflectance would be needed to measure potential para-hydrogen disequilibrium and the circulation it implies. A better understanding of the dynamics and seasonal changes taking place will require continued spatially resolved observations over seasonal time-scales (i.e. decades) and at a range of wavelengths necessary to resolve ambiguities in the observed reflectance. Even so, the great distance and extreme axial tilt of Uranus limit our views from Earth. A designated spacecraft mission will ultimately be necessary in order to make significant advances in our understanding of the atmosphere and how it may change in time.

**Acknowledgements:**

The authors wish to acknowledge the contributions of Barney Conrath, who graciously shared his time and remarkable expertise through discussions of the retrieval algorithm. We thank Daphne Stam for adapting hydrogen absorption coefficients from Borysow. We wish to acknowledge additional observers Phil Nicholson, Daphne Stam, and Barney Conrath, as well as the helpful staff of the Palomar Observatory. This research was initially completed while the primary author was a graduate student at Cornell University, and later revised with the patient support of Emily Rauscher at University of Michigan. Finally we thank the two anonymous reviewers for their incisive and thorough comments. Funding was made possible by a NASA Planetary Astronomy grant NNG05GH57G (observations) and a NASA Outer Planet Research grant NNX09AU25G (analysis).